\documentclass[12pt]{article}%
\usepackage{amsfonts,amsmath,amssymb,amsbsy,color,epsfig,harvard,multirow,pifont,rotating,threeparttable}
\citationstyle{dcu}
\newcommand{\cn}{\citeasnoun}

\usepackage{accents}
\usepackage{graphicx}
\usepackage{amsmath}
\usepackage{amssymb}
\usepackage{amsfonts}
\usepackage{tabu}
\usepackage{setspace}
\usepackage{threeparttable}
\usepackage{multirow}
\usepackage[capposition=top]{floatrow}
\usepackage{color, colortbl}
\usepackage{lscape}
\usepackage{caption}
\usepackage{subcaption}
\usepackage[affil-it]{authblk}
\usepackage{placeins}
\usepackage{setspace}
\usepackage{footmisc}
\usepackage{tabularx}
\usepackage{longtable}
\usepackage{enumitem,xkeyval}
\usepackage{color,soul}
\usepackage{etoolbox}
\usepackage{refcount}
\usepackage{lscape}
\usepackage{appendix}
\usepackage{hyperref}
\usepackage{cleveref}
\usepackage{refcount}
\usepackage{varwidth}
\usepackage{authblk}
\usepackage{multirow}

\providecommand{\U}[1]{\protect\rule{.1in}{.1in}}
\makeatother

\renewcommand{\theequation}{\thesection.\arabic{equation}}

\newtheorem{theorem}{Theorem}

\newtheorem{lemma}[theorem]{Lemma}
\newtheorem{assumption}{Assumption}

\newtheorem{proposition}[theorem]{Proposition}

\renewcommand{\theequation}{\thesection.\arabic{equation}}
\newcolumntype{L}{>{\centering\arraybackslash}m{2.5cm}}
\newcolumntype{K}{>{\centering\arraybackslash}m{3.5cm}}

\begin{document}
\begin{titlepage}
\thispagestyle{empty}
\title{\textbf{Semiparametric Quantile Models for Ascending Auctions with Asymmetric Bidders}\vspace{0.5in}}
\bigskip
%\author[1]{Jayeeta Bhattacharya } 
%\author[2]{Nathalie Gimenes }
%\author[1]{Emmanuel Guerre}
%\affil[1]{School of Economics and Finance, Queen Mary, University of London, UK.}
%\affil[2]{Department of Economics, PUC-Rio, Rio de Janeiro, Brazil.}
\author{Jayeeta Bhattacharya \authorcr University of Southampton, UK \and Nathalie Gimenes \authorcr PUC-Rio, Rio de Janeiro, Brazil \and Emmanuel Guerre \authorcr Queen Mary University of London, UK\vspace{1in}}
%%\affil{School of Economics and Finance, Queen Mary, University of London, UK.}
%%\author{Nathalie Gimenes}
%%\affil{Department of Economics, PUC-Rio, Rio de Janeiro, Brazil.}
%%\author{Emmanuel Guerre}
%%\affil{School of Economics and Finance, Queen Mary, University of London, UK.}
\date{September 2020}
\setcounter{Maxaffil}{0}
\renewcommand\Affilfont{\itshape\small}
\clearpage\maketitle
\thispagestyle{empty}
\end{titlepage}
\newpage
\thispagestyle{empty}
\begin{center}
\textbf{Abstract}
\end{center}

The paper proposes a parsimonious and flexible semiparametric quantile regression specification for asymmetric bidders within the independent private value framework. Asymmetry is parameterized using powers of a parent private value distribution, which is generated by a quantile regression specification. As noted in \cn{Cantillon2008}, this covers and extends models used for efficient collusion, joint bidding and mergers among homogeneous bidders. The specification can be estimated for ascending auctions using the winning bids and the winner's identity. The estimation is in two stage. The asymmetry parameters are estimated from the winner's identity using a simple maximum likelihood procedure. The parent quantile regression specification can be estimated using simple modifications of \cn{Gimenes2017}. Specification testing procedures are also considered. A timber application reveals that 
weaker bidders have $30\%$ less chances to win the auction than stronger ones.
It is also found that increasing participation in an asymmetric ascending auction may not be as beneficial as using an optimal reserve price as would have been expected from a result of \cn{BulowKlemperer1996} valid under symmetry.

\bigskip

\bigskip

\bigskip


\bigskip

\bigskip

{\footnotesize Nathalie Gimenes and Emmanuel Guerre acknowledge the British Academy and Newton Fund for
generously funding this project (reference code: AF150085). Nathalie Gimenes thanks Ying Fan and Ginger Jin for very useful comments. Many comments from seminar and conference participants have helped  to improve the paper.}

\newpage

\setcounter{page}{1}

\section{Introduction}

\bigskip

Asymmetry among bidders may arise from many factors, for example, differences in taste or specialization, degree of information, productivity, costs, firm size, joint bidding or collusion among a subgroup of buyers. It is, therefore, likely that symmetric bidding is only a theoretical approximation that may not fit well many auction markets.  
Within the independent private value paradigm (IPV hereafter), the revenue equivalence theorem no longer holds
with asymmetric bidders and first-price auction can be inefficient,  see \cn{Krishna2009} and the references therein.
\cn{Cantillon2008} supports the common belief that competition is reduced by bidders’ asymmetries. She shows that asymmetry decreases the seller expected revenue in first-price and second-price auctions, when compared to revenues achieved with a benchmark symmetric private value distribution. The timber auction revenue analysis of \cn{RobertsSweeting2016} shows that reducing the participation of strong bidders can considerably lower the seller expected revenue. 

\cn{Myerson1981} suggests to depart from standard formats and describes an optimal auction which restores some competition by handicapping strong bidders. This mechanism critically involves the private value distribution and is difficult to implement. In an empirical study of snow removal contract sealed procurements, \cn{FlambardPerrigne2006} considered this optimal auction and an alternative subsidy policy. \cn{KrasnokutskayaSeim2011} studied a bid preference program for California highway auction, see also \cn{Marion2007}. \cn{AtheyCoeyLevin2013} focused on set-asides and subsidies for timber auctions.

Among the aforementioned empirical works, the only papers adopting a nonparametric approach are \cn{FlambardPerrigne2006} and \cn{Marion2007}, who studied first-price auctions. For first-price auctions, \cn{KrasnokutskayaSeim2011}, \cn{AtheyLevinSeira2011} and \cn{AtheyCoeyLevin2013} all considered parametric specifications, as did \cn{RobertsSweeting2016} for ascending auctions. 

There are, however, some works devoted to the nonparametric approach for ascending auctions with asymmetric bidders.  A theoretical nonparametric identification result with a finite number of asymmetric types, due to \cn{Komarova2013a}, shows that the asymmetric valuation distributions can be recovered from the winning bid and the identity of the winner under IPV, see also \cn{AtheyHaile2002}. \cn{BendstrupPaarsch2006} have proposed a related semi-nonparametric estimation procedure. \cn{Lamy2012} shows that nonparametric identification still holds under anonymity for second-price auctions when all the bids are observed. Set identification results are also available for affiliated models, which are not point identified as shown by \cn{AtheyHaile2002}. For affiliated values and second-price auction, \cn{Komarova2013b} gives bounds for joint private value distribution, assuming identities are available. \cn{CoeyLarsenSweeneyWaisman2017} consider a more difficult scenario, where only the winning bid is observed and anonymity is possible. They obtain bounds for the seller expected revenue and bidder surplus which extends upon the ones of \cn{AradillasGandhiQuint2013} for the symmetric case.

Developing nonparametric approaches for asymmetric bidders with a discrete number of types is difficult, because a different value distribution must be estimated for each types, as in \cn{FlambardPerrigne2006} or \cn{BendstrupPaarsch2006}. Dividing the sample in subsamples defined by a given type may result in small subsamples in addition to poor nonparametric estimation rates due to the curse of dimensionality. Comparing the valuation distribution across types is not an easy task. In this paper, we tackle these two issues through a semiparametric approach allowing for a nonparametric component common to each type and using a parametric description of type heterogeneity. The common nonparametric component is a parent private value conditional distribution $F(v|x)$, where $x$ is an auction product-specific covariate. Following \cn{Gimenes2017}, we assume that $F(v|x)$ corresponds to a quantile regression model, so that this rich and flexible specification can be estimated with a standard parametric rate independently of the dimension of $x$. The asymmetry parameter, say $\lambda_i$, is an exponent specific to bidder $i$, whose private value distribution is
\[
F_i (v|x)
=
F^{\lambda_i}
(v|x).
\]
The exponent $\lambda_i$ can be an individual fixed effect which captures unobserved bidder characteristics. As developed in the paper, it can also be a parametric function of some observed bidder variables and fixed effect parameters. In our timber application, the buyers are either mill or logger, which are considered as weak and strong bidders respectively in all applications. 

\cn{Cantillon2008} has used a similar specification for theoretical illustration purpose, noting that it has been used to ``model efficient collusion, joint bidding and mergers among homogeneous bidders'', which can be relevant for many applications. Indeed, when $\lambda_i$ is an integer number, $F^{\lambda_i} (v|x)$ is the distribution of the maximum value of $\lambda_i$ symmetric  bidders with independent valuations drawn from $F(v|x)$, as relevant, for instance, in joint bidding. This feature also shows that the asymmetry parameter $\lambda_i$ is a measure of the ``strength'' of bidder $i$. A small numerical experiment in the paper parallels \cn{Cantillon2008}, adopting an econometric point of view based on the symmetric private value distribution which would be estimated ignoring asymmetry by the quantile procedure of \cn{Gimenes2017}. Such a misspecification may lead to underestimation of the optimal reserve price and seller expected revenue.

The proposed estimation is in two stage, based upon the winning bid and identity of the winner. The first stage estimates the parameters appearing in the asymmetry exponent $\lambda_i$ using a maximum likelihood procedure based upon the winner identity. The intuition behind this procedure is that the distribution of the winner identity only depends upon the relative buyers' strength, and hence on asymmetry parameter $\lambda_i$ and not upon the common parent distribution $F(v|x)$. The second stage estimates the quantile regression specification associated with $F(v|x)$. As in \cn{Gimenes2017}, it is based on a quantile regression estimation and uses individual transformations of quantile levels, which must be estimated under asymmetry. Accounting for asymmetry leads to considering a transformation which depends upon the estimated asymmetry parameter. This latter step parallels \cn{ArellanoBonhomme1207}, who similarly estimate a quantile level transformation in a three stage quantile regression procedure.

The
empirical application illustrates the methodology using USFS timber ascending auctions. Two kinds of firms are competing: firms with manufacturing
capacity (mills, usually considered as strong bidders in the literature) and firms lacking manufacturing capabilities (loggers). We take advantage of recent advances in the quantile-regression specification literature to illustrate how well the proposed model fits the data. The estimated asymmetry exponent of the loggers is $30\%$ less than the one of the mills, suggesting that, roughly speaking, two mills should be replaced by three loggers to generate an ascending auction with similar features. The empirical application also studies the seller expected revenue as a function of the proportion of loggers and the number of buyers. It reveals economically significant variations, in the range of $9\%-20\%$ between ascending auctions attended only by loggers or only by mills. In small auctions with two bidders, changing a logger by a mill can increase the seller optimal expected revenue by $5\%$ in some cases, and still as high as $1\%$ with 12 bidders. This suggests that seller expected revenue bounds that do not account for the proportion of each type can be considerably large, and that the ones averaging over types participation, as in \cn{CoeyLarsenSweeneyWaisman2017}, can be less informative. Another finding relates to an important result of \cn{BulowKlemperer1996} stating that, under symmetry, increasing participation is more beneficial than using an optimal auction. Several violations of this result are observed, especially due to the presence of weak bidders.

The paper is organized as follows. Section \ref{quantilespecsect} presents the auction setup and the asymmetric quantile specification. Sections \ref{identsect} and \ref{sect23} give the identification strategy and discuss identification of the parameter of asymmetry under several specifications. Section \ref{revandmisspecsect} shows how to design the optimal reserve price policy when bidders are asymmetric and studies the consequences of a symmetric misspecification for the seller's expected revenue. The two-step estimator is proposed in Section \ref{EAIsect} and its asymptotic distribution is obtained. A simulation of the methodology is given in Section \ref{simsect} and an empirical application using USFS timber ascending auctions is studied in Section \ref{applsect}. The proofs of all the results given in the paper are grouped in the Appendix A. Appendix B details the test procedures used in the application. Appendix C contains tables not displayed in the application section to save space.

\section{Semiparametric quantile specifications} \label{quantilespecsect}

A single and indivisible object with observed characteristics $x\in\mathbb{R}^{d}$
is auctioned to $N\geq2$ bidders through an ascending auction. Each bidder
has a specific characteristic $Z_{i}$, $i=1,\ldots,N$. The auction
covariates $x$, the number of bidders $N$ participating in the auction
and the associated bidder covariates $Z_{i}$, $i=1,\ldots,N$ are common
knowledge to buyers and sellers, and observed by the analyst. Within the IPV
paradigm, each bidder $i=1,\cdots,N$ is assumed to have a private value
$V_{i}$ for the auctioned good, which is not observed by other bidders. The
bidder only knows his own private value, but it is common knowledge for
bidders and sellers that each private value has been independently drawn from
a c.d.f. $F_{i}(\cdot|X,Z_{i})$ conditional upon $\left(X,Z_{i}\right)$, $X = \left(1,x^{\prime}\right)^{\prime}$, or equivalently,
with a conditional quantile function%
\begin{equation}
V_{i}(\tau|X,Z_{i}):=F_{i}^{-1}(\tau|X,Z_{i}),\quad\tau\text{ in }\left[
0,1\right]  \text{.}\label{Qrep}%
\end{equation}
It will be assumed later on that the analyst observes $L$ identically drawn
auctions. For each auction $\ell$, the winning bid $W_{\ell}$ and
winner's identity, the number $N_{\ell}$ of bidders, the  product-specific covariate $X_{\ell}$ and
the bidder characteristics $Z_{\ell}=\left[  Z_{1\ell},\ldots,Z_{N_{\ell} \ell
}\right]  $ are observed. As shown later,
the assumption that the identity of the winner is observed can be relaxed when
bidders are characterized using discrete types. In this case, it is sufficient
to observe the type of the winner and the numbers of bidders within a given type.

As in the symmetric private value setting, the dominant strategy for non-winners is to bid up to their true valuation. It will, therefore, be assumed that

\begin{assumption}
\label{W} The winning bid is the second-highest bidder's private value.
\end{assumption}

See  \cn{BendstrupPaarsch2006}, \cn{AradillasGandhiQuint2013}, \cn{CoeyLarsenSweeneyWaisman2017}, and \cn{Gimenes2017} for similar assumptions and
related discussions, and \cn{HaileTamer2003} for a more general incomplete game framework.

\subsection{Asymmetric private value quantile specification}

The proposed model combines an  asymmetry function known up to parameters
\begin{equation}
\lambda_{i}=\lambda(Z_{i};\alpha_{i},\beta)>0\label{lambdaspec}%
\end{equation}
with a parent conditional distribution $F\left(  \cdot|X\right)  $ which only
depends upon the product-specific covariates and is generated by a quantile-regression model\footnote{Our approach carries over with minor modifications for other quantile semiparametric specifications, such as the exponential one
$
F^{-1}\left(  \tau|X\right)  
=
\exp
\left(
X^{\prime}\gamma\left(  \tau\right)
\right)
$.
}
\begin{equation}
F^{-1}\left(  \tau|X\right)  =X^{\prime}\gamma\left(  \tau\right)
\label{Fspec}%
\end{equation}
assuming that the first entry of $X$ is a constant term. In (\ref{lambdaspec}%
), the $\alpha_{i}$ are bidder fixed effects parameter which can capture some
unobserved bidder heterogeneity. In what follows $\alpha=\left[  \alpha
_{1},\ldots,\alpha_{N}\right]  $.

The quantile regression specification (\ref{Fspec}) can be interpreted as follows. In the symmetric case where the private values $V_i$ are drawn from the parent distribution, the random quantile level $U_i=F(V_i|X)$, which indicates the rank of bidder $i$ in the private value distribution, is a measure of efficiency. As $V_i=X^{\prime}\gamma(U_i)$, the quantile regression model postulates an additive but linear contribution of the auction characteristics to bidder $i$ private value.  This contribution is summarized by the slope coefficient $\gamma(U_i)$, which does not need to be constant in most of its components as it would be for a regression model. This adds some flexibility and was found useful in our application. Under asymmetry, it holds $V_i=X^{\prime}\gamma\left[U_i^{1/\lambda_i (Z_{i};\alpha_{i},\beta)}\right]$by (\ref{Vi_spec}) below, so that a larger asymmetry coefficient $\lambda_i (Z_{i};\alpha_{i},\beta)$ gives a $U_i^{1/\lambda_i (Z_{i};\alpha_{i},\beta)}$ closer to $1$, increasing  the private value for a given efficiency $U_i$. 

\begin{assumption}
\label{Fi} Suppose (\ref{lambdaspec}) and (\ref{Fspec}) hold. There are some
$\alpha$, $\beta$ and a vector function $\gamma\left(  \cdot\right)  $ such
that%
\begin{equation}
F_{i}\left(  \cdot|X,Z_{i}\right)  =\left[  F\left(  \cdot|X\right)  \right]
^{\lambda_{i}} \label{Fispec}
\end{equation}
for all admissible $X,Z_{i}$ and all $i=1,\ldots,N$.
\end{assumption}
\cn{Cantillon2008} refers to distributions of the type of (\ref{Fispec}) as a class of distributions for which a quasi-ordering of potential bidders is available. This specification accommodates asymmetries that arise from merger, joint bidding or collusion among homogeneous bidders. See e.g. \cn{GrahamMarshall1987}, \cn{MailathZemsky1991}, \cn{McafeeMcmillan1992}, \cn{BrannmanFroeb2000} and \cn{WaehrerPerry2003}.

Assumption \ref{Fi} is equivalent to the quantile specification%
\begin{equation}
V_{i}\left(  \tau|X,Z_{i}\right)  =X^{\prime}\gamma\left[  \tau^{1/\lambda
(Z_{i};\alpha_{i},\beta)}\right],  \label{Vi_spec}%
\end{equation}
which shows that asymmetry comes from a bidder specific transformation of the
quantile level $\tau$. As detailed below, the power specification is
particularly convenient to establish identification. Examples of 
$\lambda(Z_{i};\alpha_{i},\beta)$ are given later on. The slope
coefficient $\gamma\left(  \cdot\right)  $ is the nonparametric element of the
model. It can, however, be estimated with a parametric rate as expected from the
quantile regression and shown later on. The asymmetric power exponent
$1/\lambda(Z_{i};\alpha_{i},\beta)$ measures the bidder strength: if
$\lambda(Z_{i};\alpha_{i},\beta)>\lambda(Z_{j};\alpha_{j},\beta)$ then bidder
$i$ dominates bidder $j$ in a first-order stochastic dominance sense,  i.e. $F_{i} (\cdot|X,Z_{i}) \leq F_{j} (\cdot | X,Z_{j})$ with a strict inequality inside the common support of these distribution. Note
that the private value distributions have the same support $\left[  V\left(
0|X\right)  ,V\left(  1|X\right)  \right]  $ of the parent distribution. When
$\lambda(Z_{i};\alpha_{i},\beta)$ goes to infinity, $V_{i}\left(  \tau
|X,Z_{i}\right)  $ converges to $V\left(  1|X\right)  $ while it goes to
$V\left(  0|X\right)  $ when $\lambda(Z_{i};\alpha_{i},\beta)$ goes to $0$.

Additional standard assumptions on the parent quantile slope function
$\gamma\left(  \cdot\right)  $ and the function $\lambda\left(  \cdot
;\cdot\right)  $ are as follows. In the last assumption, $\Theta$ is the
compact set of admissible asymmetry parameters $\left(  \alpha,\beta\right)  $
and $\mathcal{Z}$ is the compact support of the bidder characteristic $Z_{i}$.

\begin{assumption}
\label{X} The vector of auction specific variables, $X=\left[ 1,x_{0}%
^{\prime}\right]  ^{\prime}$, has a dimension of $\left(  d+1\right)  \times 1$. The random vector $x_{0}$ has a compact support $\mathcal{X}_{0}\subset\left(
0,+\infty\right)  ^{d}$. The matrix $\mathbb{E}\left[  XX^{\prime}\right]  $
has an inverse.
\end{assumption}

\begin{assumption}
\label{V} $V\left(  \cdot|X\right)  $ is continuously differentiable over $(0,1)$ with a
derivative $V^{\left(  1\right)  }\left(  \cdot|X\right)  $ which is strictly
positive for all $X$ in $\mathcal{X=}\left\{  1\right\}  \times\mathcal{X}%
_{0}$.
\end{assumption}

\begin{assumption}
\label{Lambda}It holds that $\inf_{\left(  z,a,b\right)  \in\mathcal{Z}\times
\Theta}\inf_{1\leq i\leq N}\lambda\left(  z;a_{i},b\right)  >0$. The function
$\lambda\left(  z;a_{i},b\right)  $ is twice continuously differentiable with
respect to $a_{i}$ and $b$. The true value $\left(  \alpha,\beta\right)  $ of
the asymmetry parameter lies in the interior of $\Theta$.
\end{assumption}

\subsection{Identification} \label{identsect}

The proposed identification procedure is in two steps, which are constructive
enough to develop a simple estimation procedure. The first step aims to identify
the bidder asymmetry parameters $\alpha$ and $\beta$ from the observed winner's identity.
Let 
$
G\left(  w|X,Z,N,i\right)
$
be the c.d.f. of the winning bid given that bidder 
$i$ wins the auction, given covariates $X$ and $Z$. Define also
\begin{equation}
\Psi_{i}\left(  \tau;Z,N,\alpha,\beta\right)  =\frac{\Lambda_{N}(Z;\alpha
	,\beta)\tau^{\Lambda_{N|i}(Z;\alpha,\beta)}-\Lambda_{N|i}(Z;\alpha,\beta
	)\tau^{\Lambda_{N}(Z;\alpha,\beta)}}{\lambda(Z_{i};\alpha_{i},\beta)}
\label{Psii}%
\end{equation}
where%
\begin{align*}
\Lambda_{N}(Z;\alpha,\beta)  &  =\sum_{j=1}^{N}\lambda(Z_{j};\alpha_{j}%
,\beta),\\
\Lambda_{N|i}(Z;\alpha,\beta)  &  =\Lambda_{N}(Z;\alpha,\beta)-\lambda
(Z_{i};\alpha_{i},\beta).
\end{align*}
The next Lemma describes the joint distribution of the winner's identity and the winning bid.

\begin{lemma}
\label{Ident1}Suppose Assumptions \ref{W} and \ref{Fi} hold. Then for any
$i=1,\ldots,N$%
\begin{eqnarray}
&
\mathbb{P}\left(  \left.  \text{Bidder }i\text{ wins the auction}\right\vert
X,Z\right)  =\frac{\lambda(Z_{i};\alpha_{i},\beta)}{\sum_{j=1}^{N}%
\lambda(Z_{j};\alpha_{j},\beta)}, \label{Abident}%
\\
&
G\left(  w|X,Z,N,i\right)  =\Psi_{i}\left[  F\left(  w|X\right)  ;Z,N,\alpha
,\beta\right]  .
\label{Gwi}
\end{eqnarray}
\end{lemma}

Suppose that the system of equations with unknowns $a$ and $b$ in $\Theta$%
\begin{equation}
\frac{\lambda(Z_{i};a_{i},b)}{\sum_{j=1}^{N}\lambda(Z_{j};a_{j},b)}%
=\frac{\lambda(Z_{i};\alpha_{i},\beta)}{\sum_{j=1}^{N}\lambda(Z_{j};\alpha
_{j},\beta)},\quad i=1,\ldots,N, \label{Eqab}%
\end{equation}
has a unique solution, $\alpha$ and $\beta$. Then, Lemma \ref{Ident1} shows
that the winner's identity distribution identifies the asymmetry parameters
$\alpha$ and $\beta$. Identification on a case by case basis with examples of functions $\lambda(\cdot;\cdot,\cdot)$ and parameter set $\Theta$ ensuring
identification of the asymmetry parameters is given in the next section. The probability of winning is very often used to assess the presence of asymmetry among the bidders, see \cn{LaffontOssardVuong1995}, \cn{FlambardPerrigne2006} for first-price sealed bid auctions and \cn{BendstrupPaarsch2006} for ascending auctions.

Identification of the parent quantile regression slope $\gamma\left(
\cdot\right)  $ follows in a second step, using the winning bid c.d.f. given
that bidder $i$ wins the auction 
in (\ref{Gwi}). The proof of Proposition 
\ref{Ident2} yields that
$\Psi_{i}\left(  \cdot;Z,\alpha,\beta\right)  $ is strictly
increasing. Therefore the conditional winning bid quantile function
$W\left(  \tau|X,Z,i\right)  $ given that $i$ wins is%
\[
W\left(  \tau|X,Z,i\right)  =F^{-1}\left[  \Psi_{i}^{-1}\left(  \tau
;Z,\alpha,\beta\right)  |X\right]  =X^{\prime}\gamma\left[  \Psi_{i}%
^{-1}\left(  \tau;Z,\alpha,\beta\right)  \right]  
\]
and then
\begin{equation}
W\left[  \Psi_{i}\left(  \tau;Z,\alpha,\beta\right)  |X,Z,i\right]
=X^{\prime}\gamma\left(  \tau\right)  . \label{Gamident}
\end{equation}
Identification of $\gamma\left(  \cdot\right)  $ easily follows as stated in
the next Proposition.

\begin{proposition}
\label{Ident2}Suppose that Assumptions \ref{W}-\ref{X} hold, and that the
asymmetry parameters $\left(  \alpha,\beta\right)  $ are identified. Then the
parent slope function $\gamma\left(  \cdot\right)  $ is also identified.
\end{proposition}

\subsection{Identified bidder asymmetry specifications} \label{sect23}

Establishing
identification of the asymmetry parameter is essential, which holds for the
following standard choice of the function $\lambda\left(  \cdot;\cdot
,\cdot\right)  $ under proper standardization of the asymmetry parameter. For
the third and fourth examples given below, it is useful to assume that the bidder covariate
$Z_{i\ell}$ varies across auctions.

\paragraph{Example 1: Bidder fixed effects.}

In this example $\lambda(Z_{i};\alpha_{i},\beta)=\alpha_{i}$, and (\ref{Eqab})
shows that asymmetry parameter identification holds provided the system of equations with unknown $a=\left[  a_{1},\ldots,a_{N}\right]  $ in $\Theta$%
\[
\frac{a_{i}}{\sum_{j=1}^{N}a_{j}}=\frac{\alpha_{i}}{\sum_{j=1}^{N}\alpha_{j}%
},\quad i=1,\ldots,N,
\]
has a unique solution. As well known, this is ensured when%
\[
\Theta=\left\{  a\in\mathbb{R}_{+\ast}^{N}|a_{1}=1\right\}
\]
that is, the parent private value distribution is the first bidder private
value distribution.\footnote{It is however possible to identify $\alpha_1$, strengthening Assumption 4 to ensure $V^{(1)} (\cdot)$ exists and is strictly positive over $[0,1]$. If so it holds for $V_1(\tau)=V(\tau^{1/\alpha_1})$, $V_1 (\tau)-V_1 (0|)$ is equivalent to $V^{(1)} (0) \tau^{1/\alpha_1}$, showing that $\alpha_1$ is identified from the lower tail of the identified parent private value distribution. Implementing this in practice may however give nonparametric consistency rates and is therefore not attempted here.}
Alternatively the simplex $\Theta=\left\{  a\in
\mathbb{R}_{+\ast}^{N}|\sum_{i=1}^{N}a_{i}=1\right\}  $ is also
possible.$\hfill\square$

\paragraph{Example 2: Linear regression.}

The case of the regression specification $\lambda(Z_{i};\alpha_{i}%
,\beta)=Z_{i}^{\prime}\beta$ is particularly useful when the covariate $Z_{i}$
codes bidder types.\footnote{Alternatively, a fixed effects specification as
in Example 1 can be used provided the fixed effects $\alpha_{i}$ can only take
$K$ unknown values $\mu_{1},\ldots,\mu_{K}$, where $K$ is the number of
types.} An example of continuous $Z_{i}$ is provided by construction procurement, where $Z_{i}$ can group the bidder's distance to the construction site and her capacity. When $\beta\neq0,$ (\ref{Eqab}) gives the system%
\[
\frac{Z_{i}^{\prime}b}{\sum_{j=1}^{N}Z_{j}b}=\frac{Z_{i}^{\prime}\beta}%
{\sum_{j=1}^{N}Z_{i}^{\prime}\beta},\quad i=1,\ldots,N,
\]
which is equivalent to $Z_{i}^{\prime}bZ_{j}^{\prime}\beta=Z_{i}^{\prime}\beta
Z_{j}^{\prime}b$ or $b^{\prime}Z_{i}Z_{j}^{\prime}\beta =\beta^{\prime}Z_{i}Z_{j}^{\prime}b $ for all bidder pair $\left\{  i,j\right\}  $. If the range of $Z_{i}Z_{j}^{\prime}$ has a non-empty interior, differentiating with respect to the entries of this matrix gives  $b_{p_1} \beta_{p_2} = \beta_{p_1} b_{p_2}$ for all pair $(p_1, p_2)$. Hence, $\beta$ is identified up to a multiplicative constant and imposing that the first entry of $\beta$ is 1 or that $\beta^{\prime}\beta=1$ ensures
identification.$\hfill\square$

\paragraph{Example 3: Linear regression with bidder fixed effects.}

The case of $\lambda(Z_{i\ell};\alpha_{i},\beta)=\alpha_{i}+Z_{i\ell
}^{\prime}\beta$ can be dealt as in Example 2, augmenting $Z_{i\ell}$ to code
bidder identities.$\hfill\square$

\paragraph{Example 4: Exponential linear regression with bidder fixed
effects.}

When the $Z_{i\ell}$ entries can take negative values, a possible choice of
the positive function $\lambda\left(  \cdot;\cdot,\cdot\right)  $ is
$\lambda(Z_{i\ell};\alpha_{i},\beta)=\alpha_{i}\exp\left(  Z_{i\ell}^{\prime
}\beta\right)  $. For this choice, taking logarithm in (\ref{Eqab}) implies
\[
\ln a_{j} - \ln a_{i}
-
\left(
\ln \alpha_{j} - \ln \alpha_{i}
\right)
+
\left(Z_{j\ell} - Z_{i \ell} \right)^{\prime}(b-\beta)
=0
,
\quad
\text{for all $1 \leq i <j \leq N$.}
\]
If $\mathrm{Var} \left( Z_{j\ell} - Z_{i \ell} \right) \neq 0$ for some pair $(i,j)$, it must hold that $b=\beta$ and then $a_j/a_i = \alpha_j/\alpha_i$ for all $1 \leq i <j \leq N$. Restricting the parameter space of the $\alpha_i$ as in Example 1 then gives identification. $\hfill\square$

\setcounter{equation}{0}

\section{Seller revenue and asymmetry misspecification} \label{revandmisspecsect}

The proposed specification is convenient to compute and analyze the seller
revenue. The presence of a reserve price $R=R\left(  X,Z,V_{0}\right)  $,
where $V_{0}$ is the seller private value, requires changing Assumption
\ref{W} into

\begin{assumption}
\label{R} There is no transaction if all private values are below the
reserve price. Otherwise, the winning bid is the greater of the
second-highest bidder's private values and the reserve price.
\end{assumption}

For a reserve price in the common support $\left[  V\left(  0|X\right)
,V\left(  1|X\right)  \right]  $, consider the quantile level $r=r\left(
X,Z,V_{0}\right)  =F\left(  R|X\right)  $ of $R$ in the parent distribution.
Under Assumption \ref{V} it therefore holds that $R=V\left(  r|X\right)  $. It
is convenient to abbreviate $\lambda\left(  Z_{i};\alpha_{i},\beta\right)  $,
$\Lambda_{N}\left(  Z;\alpha,\beta\right)  $, $\Lambda_{N|i}\left(
Z;\alpha,\beta\right)  $ into $\lambda_{i}$, $\Lambda_{N}$ and $\Lambda_{N|i}$,
respectively. The seller payoff in an auction with reserve price $R$ is%
\[
\pi\left(  r\right)  =W\mathbb{I}\left(  W\geq R\right)  +V_{0}\mathbb{I}%
\left(  W<R\right),  
\]
where $W$ is the winning bid.
The corresponding expected seller revenue is%
\[
\Pi\left(  r|X,Z,V_{0}\right)  =\mathbb{E}\left[  \pi\left(  r\right)
|X,Z,V_{0}\right]   .
\]

\subsection{Expected revenue and optimal reserve price}

The next Proposition gives a quantile expression for the expected revenue and
characterizes the optimal reserve price. Let $\Lambda_{N} = \Lambda_{N}(Z;\alpha,\beta)$  and $\Lambda_{N | i} = \Lambda_{N|i}(Z;\alpha,\beta)$ be as (\ref{Psii}).

\begin{proposition}
\label{ERRP} Suppose Assumptions \ref{Fi}, \ref{V} and \ref{R} hold. Then

\begin{enumerate}
\item[(i)] The probability of selling the auctioned good is $(1-r^{\Lambda
_{N}})$.

\item[(ii)] The seller expected payoff is
\begin{align}
\Pi\left(  r|X,Z,V_{0}\right)   &  =V_{0}r^{\Lambda_{N}}+R\sum_{i=1}%
^{N}r^{\Lambda_{N|i}}\left(  1-r^{\lambda_{i}}\right) \nonumber\\
&  +\int_{r}^{1}V\left(  t|X\right)  \left\{  \left(  1-N\right)  \Lambda
_{N}t^{\Lambda_{N}-1}+\sum_{i=1}^{N}\Lambda_{N|i}t^{\Lambda_{N|i}-1}\right\}
dt. \label{ER}%
\end{align}

\item[(iii)] The optimal reserve price $R_{\ast}=V\left(  r_{\ast
}|X\right)  $ satisfies
\begin{equation}
V_{0}=R_{\ast}-V^{(1)}(r_{\ast}|X)\frac{r_{\ast}}{\Lambda_{N}}\sum
_{i=1}^{N}\left(  r_{\ast}^{-\lambda_{i}}-1\right)  . \label{RP}%
\end{equation}

\end{enumerate}
\end{proposition}

Compared to the case of symmetric bidders, Proposition \ref{ERRP}-(ii) shows
that the optimal reserve price depends upon the number $N$ of bidders and
upon the bidder characteristics.  The impact of the
asymmetry coefficients $\lambda_{i}$ on the expected seller revenue and on the
optimal reserve price seems unclear. For $\Pi\left(  r\right)  $, the ambiguity is due to the term $-r^{\Lambda_{N|i}+\lambda_{i}}$ which increases with $\lambda_{i}$, while the other terms decrease. Observe similarly that, in (\ref{RP}), $1/\Lambda_{N}$
decreases with $\lambda_{i}$ while $r^{-\lambda_{i}}$ increases. Cantillon
(2008, Theorem 2) gives condition that allows to rank two sets of asymmetry
coefficients $\lambda_{i}$ according to seller revenue.

In many cases, the seller must decide a reserve price before observing the number $N$ of bidders and the asymmetry parameter of the entrants. The expected revenue formula (\ref{ER}) is conditional on $N$ and on the asymmetry parameters of the entrant, an information which is not available but can be integrated out to produce a relevant expected revenue and optimal reserve price.  

\subsection{The effect of a symmetric misspecification \label{Symmiss}}

To analyse the effect of a symmetric misspecification on the optimal
reserve price and seller revenue, we perform a numerical experiment with
no covariate and two asymmetric bidders with private values $F_{i}\left(
v\right)  =\left(  v^{\kappa}\right)  ^{\lambda_{i}}$, $0\leq v\leq1$ and
$i=1,2$. Higher $\kappa$ gives private values closer to 1 and values of the
curvature parameter $\kappa$ ranging from $1$ to $50$ are considered. High and
moderate asymmetry scenarios, with $\left(  \lambda_{1},\lambda_{2}\right)  $
set to $\left(  0.1,3.9\right)  $ and $\left(  0.1,0.9\right)  $ respectively
are considered.

To evaluate the effect of estimating a symmetric misspecified model, we derive the limiting symmetric private value distribution by matching the distribution of winning bids with the symmetric winning bid distribution. Under Assumption \ref{W}, the winning bid is equal to the minimum between $(V_{1}
,V_{2})$, therefore, the
winning bid distribution is
\[
F_{\lambda,W}\left(  w\right)  =\mathbb{P}\left(  \min\left(  V_{1}%
,V_{2}\right)  \leq w\right)  =w^{\kappa\lambda_{1}}+w^{\kappa\lambda_{2}%
}-w^{\kappa\left(  \lambda_{1}+\lambda_{2}\right)  },\quad w\in\left[
0,1\right]  .
\]
For symmetric bidders, the function $\Psi_{i}\left(  \cdot\right)  $ does not
depend upon $i$ and is equal to
\[
\Psi\left(  \tau\right)  =2\tau-\tau^{2}=1-\left(  1-\tau\right)  ^{2}.
\]
Therefore, the symmetric private value c.d.f. $F_{\lambda,S}\left(
\cdot\right)  $ which generates the winning bid distribution $F_{\lambda
,W}\left(  \cdot\right)  $ must satisfy $\Psi\left[  F_{\lambda,S}\right]
=F_{\lambda,W}$ so that%
\[
F_{\lambda,S}\left(  v\right)  =1-\left(  1-v^{\kappa\lambda_{1}}%
-v^{\kappa\lambda_{2}}+v^{\kappa\left(  \lambda_{1}+\lambda_{2}\right)
}\right)  ^{1/2},\quad v\in\left[  0,1\right]  .
\]
The c.d.f. $F_{\lambda,S}\left(  v\right)  $ is the limit of any nonparametric
estimator obtained by matching the winning bid distribution of a misspecified
symmetric bidder model with the observed one, see for instance \cn{Gimenes2017}.
An optimal reserve price assuming symmetric bidders, $R_{\lambda
,S}=V_{\lambda,S}\left(  r_{\lambda,S}\right)  $ where $V_{\lambda,S}\left(
\cdot\right)  =F_{\lambda,S}^{-1}\left(  \cdot\right)  $, solves the symmetric
version of (\ref{RP})%
\[
R_{\lambda,S}-V_{\lambda,S}^{\left(  1\right)  }\left(  r_{\lambda,S}\right)
\left(  1-r_{\lambda,S}\right)  =0
\]
where the seller private value $V_{0}$ is set to $0$ for the sake of
simplicity. The expected seller revenue achieved with
$R_{\lambda,S}$ under the true asymmetric private value distribution can then
be computed using (\ref{ER}). The reserve price $R_{\lambda,S}$ and the
corresponding expected seller revenue are reported in the columns labeled
\textquotedblleft Misspecified\textquotedblright\ of the next table. The
optimal reserve price and seller revenue using the true private value
distribution are reported under the label \textquotedblleft
Asymmetry\textquotedblright.

\begin{table}[ptbh]
\caption{Misspecified symmetric versus true asymmetric models}%
\label{tab:TruevsConsEsti}
\begin{center}
\begin{tabularx}{\textwidth}{|c|c|c|c|c|c|c|c|}
			\hline\hline
			\multicolumn{1} {|X|} { } & \multicolumn{1} {|X|} { } & \multicolumn{1} {|X|} { } & \multicolumn{2} {c|} {Optimal Reserve Price} & \multicolumn{3} {c|} {Expected Seller Revenue} \\ [0.5 ex]
			\hline
			$\lambda_1$&$\lambda_2$&$\kappa$&Asymmetry&Misspecified&Asymmetry&Misspecified&{Percentage Loss}\\
			\hline
			$0.1$		&$3.9$		&$1$		&$0.6630$		&$0.5451$		&$0.5389$		&$0.5059$		&$6.12\%$\\
			$  $		&$   $		&$2$		&$0.7550$		&$0.5995$		&$0.6800$		&$0.6054$		&$10.97\%$\\
			$  $		&$   $		&$5$		&$0.8558$		&$0.6403$		&$0.8223$		&$0.6738$		&$18.06\%$\\
			$  $		&$   $		&$10$		&$0.9092$		&$0.6671$		&$0.8927$		&$0.7230$		&$19\%$\\
			$  $		&$   $		&$50$		&$0.9730$		&$0.7785$		&$0.9707$		&$0.7173$		&$26.10\%$\\
			\hline
			$0.1$		&$0.9$		&$1$		&$0.4830$		&$0.4420$		&$0.2550$		&$0.2535$		&$0.59\%$\\
			$  $		&$   $		&$2$		&$0.5559$		&$0.4901$		&$0.3948$		&$0.3887$		&$1.55\%$\\
			$  $		&$   $		&$5$		&$0.6768$		&$0.5773$		&$0.5987$		&$0.5767$		&$3.67\%$\\
			$  $		&$   $		&$10$		&$0.7676$		&$0.6450$		&$0.7336$		&$0.6930$		&$5.53\%$\\
			$  $		&$   $		&$50$		&$0.8710$		&$0.7785$		&$0.9283$		&$0.7148$		&$23\%$\\
			\hline
		\end{tabularx}
\end{center}
\end{table}

Table \ref{tab:TruevsConsEsti} reveals that ignoring asymmetry can lead to substantial loss in terms of seller revenue, when the curvature parameter $\kappa$ is high or in the high asymmetry scenario. Note that the optimal expected revenue computed under the misspecified symmetric model is always smaller than the one achieved with the correct asymmetric model. The optimal reserve price is also substantially higher in the correct model with strong asymmetry.

The analysis presented here differs from \cn{Cantillon2008}'s, who studies how presence of asymmetry impacts seller expected revenue. She compares auctions with asymmetric private value distributions with a symmetric benchmark, and finds that asymmetry is associated with lower expected revenue. Our comparison differs from hers in that we investigate what happens if a truly asymmetric model is \textit{misspecified} as symmetric - in which case, we find that not accounting for asymmetry may lead to loss in revenue. In that sense, our analysis supports and extends her findings. To see this, consider different scenarios of asymmetry such that $\lambda_1+\lambda_2=1$,  with $\kappa=1$, as shown in Table \ref{tab:VaryAsymm}.

\begin{table}[htbp]
        \begin{center}
        \caption{Varying asymmetry levels}
        \begin{tabularx}{\textwidth}{|c|c|c|c|c|c|c|}
  \hline\hline
   \multicolumn{1} {|X|} { } & \multicolumn{1} {|X|} { } & \multicolumn{2} {c|} {Optimal Reserve Price} & \multicolumn{3} {c|} {Expected Seller Revenue} \\ [0.5 ex]
  \hline
  $\lambda_1$&$\lambda_2$&Asymmetry&Misspecified&Asymmetry&Misspecified&{Percentage Loss}\\
  \hline
	$0.1$		&$0.9$		&$0.4830$		&$0.4420$		&$0.2550$		&$0.2535$		&$0.59\%$\\
	$0.2$		&$0.8$		&$0.4680$		&$0.4433$		&$0.2593$		&$0.2590$		&$0.14\%$\\
	$0.3$		&$0.7$		&$0.4550$		&$0.4442$		&$0.2627$		&$0.2627$		&$0.003\%$\\
	$0.4$		&$0.6$		&$0.4470$		&$0.4440$		&$0.2648$		&$0.2648$		&$0.0003\%$\\
	$0.5$		&$0.5$		&$0.4440$		&$0.4449$		&$0.2655$		&$0.2655$		&$0.00 \%$\\
    \hline
  \end{tabularx}
  \label{tab:VaryAsymm}
  \end{center}
\end{table}

As can be seen, higher asymmetry when $(\lambda_1, \lambda_2)$ is (0.1,0.9) has lesser revenue than the symmetric case of (0.5,0.5), supporting Cantillon's finding that ``the expected revenue is lower the more asymmetric bidders are". However, given ex-ante asymmetry among bidders, a misspecified symmetric model always has smaller expected revenue and higher the asymmetry, more is the potential loss in revenue due to misspecification. 

\setcounter{equation}{0}

\section{Estimation and asymptotic inference} \label{EAIsect}

Suppose that for each auction $\ell$, the analyst observes the winning bid
$W_{\ell}$, the product-specific covariate $X_{\ell}$, the number of bidders $N_{\ell}$,
the bidder covariate $Z_{\ell}$ and the identity $I_{\ell}^{\ast}$ of the
winner.  

\subsection{Two step estimation}

As stated in Lemma \ref{Ident1}, the probability
that bidder $i$ wins is%
\[
P\left(  i|Z_{\ell},N_{\ell},\alpha,\beta\right)  =\frac{\lambda(Z_{i\ell
};\alpha_{i},\beta)}{\sum_{j=1}^{N_{\ell}}\lambda(Z_{j\ell};\alpha_{j},\beta)}%
\]
so that the asymmetry parameter $\left(  \alpha,\beta\right)  $ can be
estimated using the maximum likelihood estimator%
\begin{equation}
\left(  \widehat{\alpha},\widehat{\beta}\right)  =\arg\max_{\left(
\alpha,\beta\right)  \in\Theta}\sum_{\ell=1}^{L}\ln P\left(  I_{\ell}^{\ast
}|Z_{\ell},N_{\ell},\alpha,\beta\right)  . \label{MLEab}%
\end{equation}
The second step consists in the estimation of the parent quantile slope and is
based upon (\ref{Gamident}), which identifies $\gamma\left(  \cdot\right)  $
as shown in Proposition \ref{Ident2}. Define, for $\Psi_{i}\left(
\tau;Z,N,\alpha,\beta\right)  $ as in (\ref{Psii}),
\[
\widehat{\Phi}_{\ell}\left(  \tau\right)  =\Phi_{\ell}\left(  \tau
;\widehat{\alpha},\widehat{\beta}\right)  =\Psi_{I_{\ell}^{\ast}}\left(
\tau;Z_{\ell},N_{\ell},\widehat{\alpha},\widehat{\beta}\right)  .
\]
The quantile level $\widehat{\Phi}_{\ell}\left(  \tau\right)  $ is an
estimation of the (random) quantile level $\Psi_{I_{\ell}^{\ast}}\left(
\tau;Z_{\ell},\alpha,\beta\right)  $ which is such that the quantile function
of the winning bid given $X_{\ell}$, $Z_{\ell}$, $N_{\ell}$ and $I_{\ell
}^{\ast}$ satisfies
\[
W\left[  \Psi_{I_{\ell}^{\ast}}\left(  \tau;Z_{\ell},\alpha,\beta\right)
|X_{\ell},Z_{\ell},N_{\ell},I_{\ell}^{\ast}\right]  =X_{\ell}^{\prime}%
\gamma\left(  \tau\right),
\]
(see (\ref{Gamident})). It suggests the quantile regression estimator%
\begin{equation}
\widehat{\gamma}\left(  \tau\right)  =\arg\min_{\gamma}\sum_{\ell=1}^{L}%
\rho_{\widehat{\Phi}_{\ell}\left(  \tau\right)  }\left(  W_{\ell}-X_{\ell
}^{\prime}\gamma\right)  \label{QRgam}%
\end{equation}
where $\rho_{\Phi}\left(  u\right)  =u\left(  \Phi-\mathbb{I}\left(
u<0\right)  \right)  $, see e.g. \cn{Koenker2005}.

\subsection{Asymptotic distribution}

While a joint estimation of the parameters $\alpha$, $\beta$ and $\gamma (\cdot)$ may offer some potential efficiency gains, the proposed two step procedure is simple to implement. In addition, the first stage estimation of $(\alpha,\beta)$ is not affected by a possible mispecification of the parent c.d.f. Note the second step slope
estimator involves an estimated quantile level. As well known since \cn{MurphyTopel1985}, the first step estimation can affect the second step asymptotic distribution, but not the rate of $\widehat{\gamma}(\cdot)$ which is still the parametric $\sqrt{L}$ rate. However, this can  be easily captured using the proof techniques in \cn{Pollard1991}. Useful assumptions and notations are as follows. In the sequel, $\left(  \alpha,\beta\right)  $ will be abbreviated in $\theta$ when convenient. Let $P^{\theta}\left(  i|Z_{\ell},N_{\ell},\theta\right)  $ be $\theta$ derivative of $P\left(  i|Z_{\ell},N_{\ell},\theta\right)  $. Under Assumption \ref{Lambda}, 
%$P^{\theta}\left(  I_{\ell}^{\ast}|Z_{\ell},N_{\ell},\theta\right)  /P\left(  I_{\ell}^{\ast}|Z_{\ell},N_{\ell},\theta\right)  $ and 
the Fisher information matrix for the asymmetry parameters can be defined as%
\[
\mathcal{I}\left(  \theta\right)  =\operatorname*{Var}\left(  \frac
{P^{\theta}\left(  I_{\ell}^{\ast}|Z_{\ell},N_{\ell},\theta\right)  }%
{P\left(  I_{\ell}^{\ast}|Z_{\ell},N_{\ell},\theta\right)  }\right)
\]
or by using the Bartlett identity when $Z_{\ell}$ has a compact support as
assumed below.

\begin{assumption}
\label{Ind} The auction variables $\left(  X_{\ell},N_{\ell},Z_{\ell},I_{\ell
}^{\ast},W_{\ell}\right)  $ are drawn identically and independently. The
support $\mathcal{Z}$ of $Z_{\ell}$ is compact.
\end{assumption}

\begin{assumption}
\label{FSident} The identification equations (\ref{Abident}) and
(\ref{Gamident}) hold. The asymmetry parameters are identified and the Fisher
information matrix $\mathcal{I}\left(  \theta\right)  $ has an inverse.
\end{assumption}

\bigskip

Assumption \ref{Ind} is standard. Assumption \ref{FSident}
imposes that the auction model is correctly specified and that the asymmetry
parameters are identified.

Consider now some additional notations for the second step estimator
$\widehat{\gamma}\left(  \tau\right)  $. Let the $\tau$ derivative of
$\Psi_{I_{\ell}^{\ast}}\left(  \tau;Z_{\ell},N_{\ell},\theta\right)  $ be denoted by $\Psi_{I_{\ell}^{\ast}}^{\tau
}\left(  \tau;Z_{\ell},N_{\ell},\theta\right)  $, which
exists and is strictly positive on $\left(  0,1\right)  $ as shown in the
proof of Proposition \ref{Ident2}. As the conditional quantile function of the
winning bid is
\[
W\left(  \tau|X_{\ell},Z_{\ell},I_{\ell}^{\ast},N_{\ell}\right)  =X_{\ell
}^{\prime}\gamma\left[  \Psi_{I_{\ell}^{\ast}}^{-1}\left(  \tau;Z_{\ell
},N_{\ell},\theta\right)  \right]
\]
the conditional p.d.f. of the winning bid is, under Assumption \ref{V},%
\begin{align*}
 f_{W}\left(  w|X_{\ell},Z_{\ell},I_{\ell}^{\ast},N_{\ell}\right)  &=\frac
{1}{W^{\left(  1\right)  }\left[  W^{-1}\left(  w|X_{\ell},Z_{\ell},I_{\ell
}^{\ast},N_{\ell}\right)  |X_{\ell},Z_{\ell},I_{\ell}^{\ast},N_{\ell}\right]
}\\
&=\frac{{\Psi_{I_{\ell}^{\ast}}^{\tau}\left(
\Psi_{I_{\ell}^{\ast}}^{-1}\left(  W^{-1}\left(  w|X_{\ell},Z_{\ell},I_{\ell
}^{\ast},N_{\ell}\right)  ;Z_{\ell},N_{\ell},\theta\right)  ;Z_{\ell},N_{\ell
},\theta\right)  }}{X_{\ell}^{\prime}\gamma^{\left(  1\right)  }\left[
\Psi_{I_{\ell}^{\ast}}^{-1}\left(  W^{-1}\left(  w|X_{\ell},Z_{\ell},I_{\ell
}^{\ast},N_{\ell}\right)  ;Z_{\ell},N_{\ell},\theta\right)  \right]  }\\
%&  \quad\quad\quad\times\frac{1}{\Psi_{I_{\ell}^{\ast}}^{\tau}\left(
%\Psi_{I_{\ell}^{\ast}}^{-1}\left(  W^{-1}\left(  w|X_{\ell},Z_{\ell},I_{\ell
%}^{\ast},N_{\ell}\right)  ;Z_{\ell},N_{\ell},\theta\right)  ;Z_{\ell},N_{\ell
%},\theta\right)  }%
\end{align*}
which is continuous and bounded away from infinity over $\left(  V\left(
0|X_{\ell}\right)  ,V\left(  1|X_{\ell}\right)  \right)  $. Let the $\theta$ derivative of $\Psi_{I_{\ell}^{\ast}}\left(  \tau;Z_{\ell
},N_{\ell},\theta\right)  $ be denoted by $\Psi
_{I_{\ell}^{\ast}}^{\theta}\left(  \tau;Z_{\ell},N_{\ell},\theta\right)  $ and define%
\begin{align*}
H\left(  \tau\right)   &  =\mathbb{E}\left[  X_{\ell}X_{\ell}^{\prime}%
f_{W}\left(  X_{\ell}^{\prime}\gamma\left(  \tau\right)  |X_{\ell},Z_{\ell
},I_{\ell}^{\ast},N_{\ell}\right)  \right]  ,\\
J\left(  \tau\right)   &  =\mathbb{E}\left[  X_{\ell}X_{\ell}^{\prime}\left(
\mathbb{I}\left(  W_{\ell}\leq X_{\ell}^{\prime}\gamma\left(  \tau\right)
\right)  -\Psi_{I_{\ell}^{\ast}}\left(  \tau;Z_{\ell},N_{\ell},\theta\right)
\right)  ^{2}\right] \\
C\left(  \tau\right)   &  =\mathbb{E}\left[ \left(X_{\ell}\left(  \mathbb{I}\left(  W_{\ell}\leq X_{\ell}^{\prime}\gamma\left(
\tau\right)  \right)  - \Psi_{I_{\ell}^{\ast}}\left(  \tau;Z_{\ell},N_{\ell
},\theta\right)  \right) \right) \mathcal{I}^{-1}\left(  \theta\right)  \left(
\frac{P^{\theta}\left(  I_{\ell}^{\ast}|Z_{\ell},N_{\ell},\theta\right)
}{P\left(  I_{\ell}^{\ast}|Z_{\ell},N_{\ell},\theta\right)  }\right)
^{\prime}\right]  ,\\
D\left(  \tau\right)   &  =-\mathbb{E}\left[  \Psi_{I_{\ell}^{\ast}}^{\theta
}\left(  \tau;Z_{\ell},N_{\ell},\theta\right)  X_{\ell}^{\prime}\right]  .
\end{align*}
The matrices $H\left(  \tau\right)  $ and $J\left(  \tau\right)  $ are
specific to the infeasible quantile regression estimator $\widetilde{\gamma
}\left(  \tau\right)  $ of $\gamma\left(  \tau\right)  $ which uses the true
asymmetry parameters $\left(  \alpha,\beta\right)  $ instead of their estimates,%
\[
\widetilde{\gamma}\left(  \tau\right)  =\arg\min_{\gamma}\sum_{\ell=1}^{L}%
\rho_{\Psi_{I_{\ell}^{\ast}}\left(  \tau;Z_{\ell},N_{\ell},\theta\right)
}\left(  W_{\ell}-X_{\ell}^{\prime}\gamma\right)  .
\]
In particular, $H^{-1}\left(  \tau\right)  J\left(  \tau\right)  H^{-1}\left(
\tau\right)  $ is the asymptotic variance of $\widetilde{\gamma}\left(
\tau\right)  $, see \cn{Koenker2005}. The matrix $C\left(  \tau\right)  $ is the
asymptotic covariance of the infeasible $\widetilde{\gamma}\left(
\tau\right)  $ and $\left(  \widehat{\alpha},\widehat{\beta}\right)  $.
Finally%
\[
D\left(  \tau\right) =\frac{\partial}{\partial\theta \partial\gamma^{\prime}}\mathbb{E}%
\left[  \rho_{\Psi_{I_{\ell}^{\ast}}\left(  \tau;Z_{\ell},N_{\ell}%
,\theta\right)  }\left(  W_{\ell}-X_{\ell}^{\prime}\gamma\left(  \tau\right)
\right)  \right]
\]
is the $\theta \gamma$ derivative of the population version of the objective function
which is used for $\widetilde{\gamma}\left(  \tau\right)  $.

The asymptotic variance of the asymmetry parameter estimator $\left(  \widehat{\alpha
},\widehat{\beta}\right)  $ and of the feasible $\widehat{\gamma}\left(  \tau\right)
$ are given by the matrices $\mathcal{I}^{-1}\left(
\theta\right)  $ and
\begin{align*}
C_{\gamma \gamma} (\tau)   &  =H^{-1}\left(  \tau\right)  \left\{  J\left(
\tau\right)  +D\left(  \tau\right)  \mathcal{I}^{-1}\left(  \theta\right)
D\left(  \tau\right)  ^{\prime}+D\left(  \tau\right)  C\left(  \tau\right)
^{\prime}+C\left(  \tau\right)  D\left(  \tau\right)  ^{\prime}\right\}
H^{-1}\left(  \tau\right) \\
C_{\gamma\theta}\left(  \tau\right)   &  =H^{-1}\left(  \tau\right)\left\{ -C\left(  \tau\right)  -D\left(
\tau\right)  \mathcal{I}^{-1}\left(  \theta\right)\right\}
\end{align*}
The next Theorem gives the asymptotic joint distribution of $\widehat{\gamma
}\left(  \tau\right)  $ and $\left(  \widehat{\alpha},\widehat{\beta}\right)
$.

\begin{theorem}
\label{AN}Suppose Assumptions \ref{Fi}-\ref{Lambda}, \ref{Ind} and
\ref{FSident} hold. Then, for any quantile level $\tau$ in $\left(
0,1\right)  $, $\widehat{\gamma}\left(  \tau\right)  $ and $\widehat{\theta
}=\left(  \widehat{\alpha}^{\prime},\widehat{\beta}^{\prime}\right)  ^{\prime
}$ are asymptotically normal with
\[
\sqrt{L}\left(  \left(  \widehat{\gamma}\left(  \tau\right)  -\gamma\left(
\tau\right)  \right)  ^{\prime},\left(  \widehat{\theta}-\theta\right)
^{\prime}\right)  ^{\prime}\overset{d}{\rightarrow}\mathcal{N}\left(  0,%
\begin{bmatrix}
C_{\gamma \gamma} (\tau)  & C_{\gamma\theta}\left(  \tau\right) \\
C_{\gamma\theta}\left(  \tau\right)  ^{\prime} & \mathcal{I}(\theta)^{-1}%
\end{bmatrix}
\right)  .
\]

\end{theorem}

While the asymptotic normality of the MLE $\widehat{\theta}$ is standard, the
one of $\widehat{\gamma}\left(  \tau\right)  $ follows from
modifying the approach of \cn{Pollard1991} to account for the first step
estimation. The asymptotic variance of these estimators can be estimated but
it may be more suitable to rely on bootstrap, especially for the parent
slope function $\gamma\left(  \cdot\right)  $. Indeed, bootstrap is more
reliable for inference in quantile regression, see \cn{Koenker2005} and the
reference therein.

\subsection{Seller revenue and optimal reserve price estimation} \label{Revenue}
Let $\widehat{\lambda}_{i}= \lambda_i (Z;\widehat{\alpha},\widehat{\beta})$, $\widehat{\Lambda}_{N} = \Lambda_{N}(Z;\widehat{\alpha},\widehat{\beta})$  and $\widehat{\Lambda}_{N | i} = \Lambda_{N|i}(Z;\widehat{\alpha},\widehat{\beta})$ be as (\ref{Psii}).
The estimated seller expected payoff derived from Proposition \ref{ERRP} is
\begin{align*}
\widehat{\Pi}\left(  r|X,Z,V_{0}\right)   
&  
=
V_{0}
r^{\widehat{\Lambda}_{N}}+
X^{\prime}
\widehat{\gamma} (r)
\sum_{i=1}^{N}
r^{\widehat{\Lambda}_{N|i}}\left(  1-r^{\widehat{\lambda}_{i}}\right) \nonumber\\
&  
+
\int_{r}^{1-\epsilon} X^{\prime}
\widehat{\gamma} (t)
\left\{  \left(  1-N\right)  
\widehat{\Lambda}_{N}t^{\widehat{\Lambda}_{N}-1}+\sum_{i=1}^{N}\widehat{\Lambda}_{N|i}t^{\widehat{\Lambda}_{N|i}-1}\right\}
dt. \label{ERest}%
\end{align*}
In the integral upper bound above, $\epsilon>0$ is a small truncation index, set to $.1$ in the Application Section, which accounts for the fact that the quantile regression estimator may not be defined for extreme quantiles. It follows that the argument $r$ must vary in $[\epsilon,1-\epsilon]$. 
As in \cn{LiPerrigneVuong2003}, an optimal reserve price estimation
$\widehat{R}_{\ast} = X^{\prime} \widehat{\gamma} \left( \widehat{r}_{\ast} \right)$  could be based on  the maximization
\[
\widehat{r}_{\ast} = \arg \max_{r \in [\epsilon,1-\epsilon]} \widehat{\Pi}\left(  r|X,Z,V_{0}\right) 
\]
instead of (\ref{RP}), as the latter would request an additional estimation of the derivative of $V(\cdot|X)$.
Note that the use of a truncation may affect the estimation of the optimal reserve price. As the values of $\widehat{r}_{\ast}$ relevant in our application were close to .5, we do not think it affects our empirical results.\footnote{Note also that the function
$t \mapsto  \left(  1-N\right)  
\widehat{\Lambda}_{N}t^{\widehat{\Lambda}_{N}-1}+\sum_{i=1}^{N}\widehat{\Lambda}_{N|i}t^{\widehat{\Lambda}_{N|i}-1}$ vanishes at $t=0$ as long as $\widehat{\Lambda}_{N}>1$, and at $1$, since $\widehat{\Lambda}_{N}= \sum_{i=1}^{N} \widehat{\lambda}_i$ and 
$\widehat{\Lambda}_{N|i}=\widehat{\Lambda}_{N}-\widehat{\lambda}_i$. This also suggests that the lower and upper tails have a moderate contribution in the expected revenue integral, at least for reasonable value of $N$.}

A Functional Central Limit Theorem can be established for $\left\{\widehat{\Pi}\left(  r|X,Z,V_{0}\right) , r \in [\epsilon,1-\epsilon] \right\}$ combining arguments used for Theorem \ref{AN} with empirical process theory as reviewed in \cn{vanderVaart1998}. The Argmax Theorem can then be used to obtain the asymptotic distribution of $\widehat{r}_{\ast}$ and of $\widehat{R}_{\ast}$. In the application, pairwise bootstrap is used to derive (pointwise) confidence intervals as proposed in \cn{Koenker2005}.

\setcounter{equation}{0}

\section{Simulations} \label{simsect}

In this section, we present the results of a Monte Carlo simulation designed
to evaluate the performance of the two-step estimation procedure.

\paragraph{Data Generating Process.} We simulate $L=2000$ ascending auctions
with $N=5$ bidders assigned to $K=2$ different classes: type 1 and type 2
with $\lambda_1=1$ and $\lambda_2=\exp{(2)}=7.39$. Bidders are assigned to each type with equal probability. Auction specific characteristics $x_{\ell}$ is a random draw from $\mathcal{U}_{\left[1,3\right]}$ , for $\ell=1 \ldots L$, with an expected value of $E[x_{\ell}] = 2$ and $X_{\ell}=[1,x_{\ell}]$. The parent private value conditional quantile function is generated as 
\begin{equation}\label{eq:MC1}
V\left(\tau|X_{\ell}\right) = X_{\ell}^{\prime}\gamma(\tau) = \gamma_{0}\left(\tau\right) + \gamma_{1}\left(\tau\right)x_{\ell},
\end{equation}
where the true quantile regression coefficients are 
\begin{align*}
\gamma_{0}\left(\tau\right) = {\tau^{\exp(1.5)}}/{2};\quad \quad \quad \gamma_{1}\left(\tau\right) = {\tau^{\exp(1.5)}}/{4}.
\end{align*}
The number of simulation replications is set to 1000.
\paragraph{Estimation.} The estimation is conducted in two steps. In the first
step, the type parameters $(\lambda_1,\lambda_2)$ are estimated using maximum likelihood estimation by maximizing \eqref{MLEab} over a grid of points. The quantile regression slope $(\gamma_0 (\tau),\gamma_1 (\tau))$ are then estimated in a second step using \eqref{QRgam}. For the median $x_{\ell}$, the estimated parent quantile function is given by $\widehat{V}\left(\tau \vert X_{\ell} \right) = \widehat{\gamma}_{0}\left(\tau\right) + 2\widehat{\gamma}_{1}\left(\tau\right)$. 

\begin{figure}[htbp]
	\centering
	\caption{Simulation: True vs. Estimated Private Value Quantile Function}
	\subcaption*{Median Auction}
	\includegraphics[width=.5\textwidth]{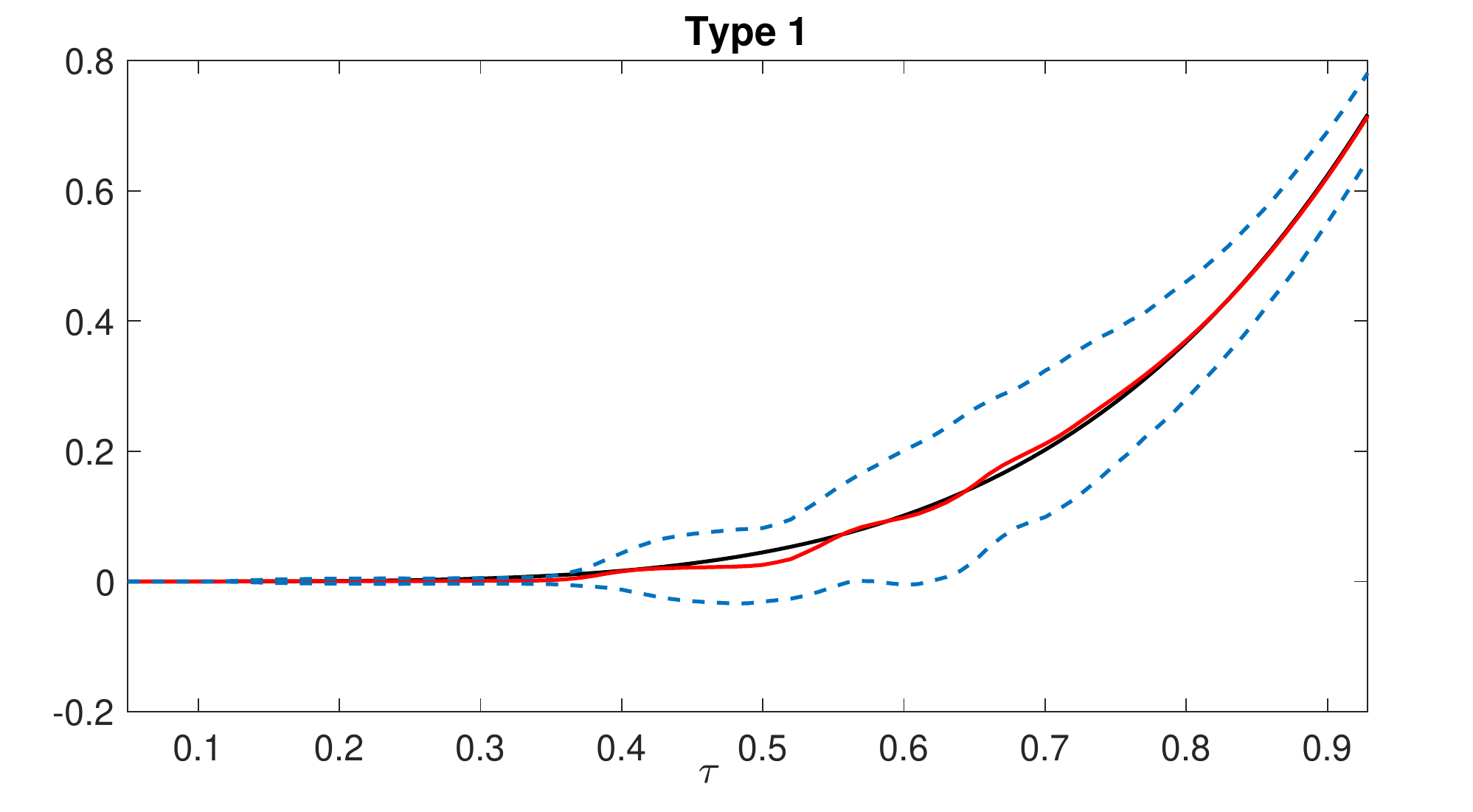}\hfill
	\includegraphics[width=.5\textwidth]{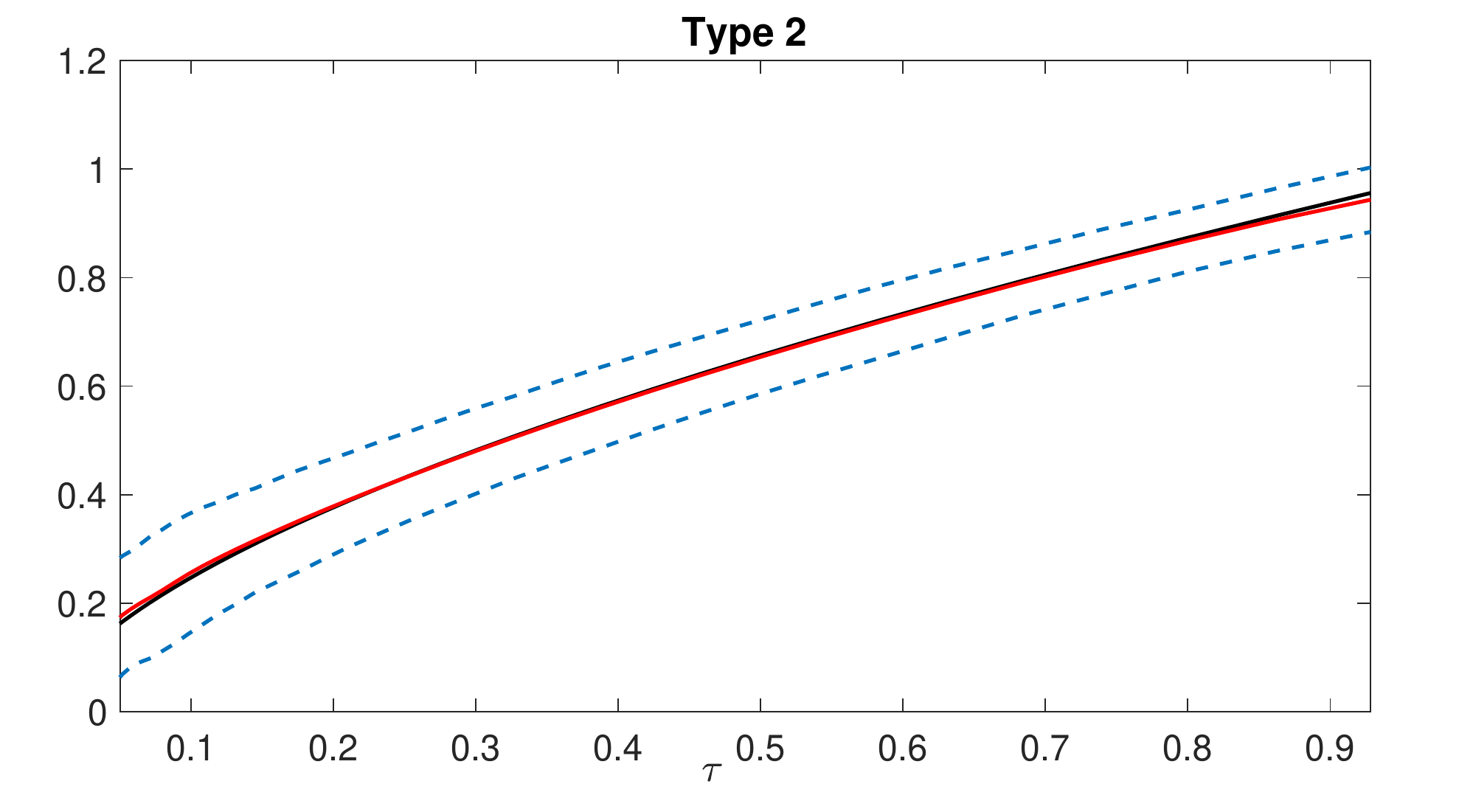}\hfill
	\label{fig:monte}
	\small\textsuperscript{True in black,  mean of estimation across simulations in red, 95\%  confidence intervals in dotted lines}
\end{figure}

\paragraph{Results.} Figure \ref{fig:monte} compares the true private value quantile function (in black) with the  mean of the estimated private value quantile function across simulations (in red) for both type $1$ and type $2$, considering a median $x_{\ell}$ auction. The bias and standard error (SE) for the private value quantile function for both types are reported in Table \ref{tab:biasSE}. The simulation results confirm that the two step estimation procedure works well.

\FloatBarrier
\FloatBarrier

\begin{table}[hptb]
\caption{Simulation: Bias and SE of Private Value Quantile Function}%
\label{tab:biasSE}
\begin{center}
\begin{tabularx}{0.7\textwidth}{ccccc}
      \hline
			\hline
			\multicolumn{1} {X} { } & \multicolumn{2} {c} {Type 1} & \multicolumn{2} {c} {Type 2} \\ [1 ex]
			\hline
			$\tau$&Bias&SE&Bias&SE\\
			\hline
			$0.1$		&$0.0000$ 	&$0.0000$		&$0.0092$	&$0.0560$		\\
			$0.2$		&$-0.0003$	  &$0.0019$		&$0.0020$ 	&$0.0460$		\\
			$0.3$		&$-0.0037$		&$0.0022$		&$-0.0013$	&$0.0401$		\\
			$0.4$		&$-0.0010$		&$0.0143$		&$-0.0023$	  &$0.0474$		\\
			$0.5$		&$-0.0192$		&$0.0288$		&$-0.0028$	  &$0.0348$		\\
			$0.6$		&$-0.0032$		&$0.0526$		&$-0.0030$	  &$0.0335$		\\
			$0.7$		&$0.0091$		&$0.0574$		&$-0.0033$ 	&$0.0309$		\\
			$0.8$		&$0.0024$		&$0.0460$		&$-0.0053$		&$0.0291$		\\
			$0.9$		&$-0.0026$		&$0.0357$		&$-0.0103$		&$0.0300$		\\
			\hline
			\hline
		\end{tabularx}
\end{center}
\end{table}

\setcounter{equation}{0}

\section{Application} \label{applsect}

In this section, we investigate asymmetry in USFS timber auctions, as reported on Phil Haile's website \texttt{http://www.econ.yale.edu/$\sim$pah29/timber/timber.htm}, using the methodology developed in this paper. Bidders are classified as mill (with manufacturing capacity to process the timber) and logger (lacking manufacturing capabilities). For simplicity in the exposition, mills and loggers are abbreviated by $M$ and $L$, respectively, when convenient. The dataset aggregates 7,462 ascending auctions (i.e., winning bids) that occurred in the western part of the US during the period of 1982-90. The sample contains a set of variables characterizing each timber tract including the estimated volume of the timber (measured in thousand of board feet - mbf) and its estimated appraisal value (given in Dollar per unit of volume). Mills won in about 72\% of the auctions.  The descriptive statistics can be found in Table \ref{desc stats}. The auctioned tract exhibits significant heterogeneity in quality and size. The contracts to extract the timber last, on average, 2 years. Bidders participation is high. On average, there are 6 bidders attending the auctions in a range of 2 to 12. 
\begin{table}[!htbp]
\centering
\caption{Descriptive Statistics}
\begin{tabular}{lcccc}
\hline\hline
  & Mean	 & Std. Dev. &	Max & Min  \\
  \hline
 Winning bids (\$ per tbf) &	126.43 &	136.22 &	5,145.71 &	0.14	\\
 Appraisal value (\$ per tbf) &	58.65	& 60.35	& 793.62 &	0.25 \\
 Volume (tbf) &	4,466.89 &	4,418.41 &	39,920 &	 8	\\
 Contract Length (years) & 1.96 & 1.3 & 42 & 0.1\\ 
 & & & &   \\
 Number of bidders &	5.77 & 	3.09 & 	12 &	 2  \\
 Number of loggers &	 1.74 &	2.10 &	11 &	0  \\
 Number of mills &	4.03 &	3.02 &	12	&	0  \\
 Bidders in the winner's class &	4.52 &	2.73 &	12 &	 1 \\ 
\hline
\hline
\label{desc stats}
\end{tabular}
\end{table}

As we do not observe individual bidders characteristics, we consider the private value quantile regression model 
\begin{equation}
V_{mill} (\tau|x)=V (\tau|x) = x^{\prime} \gamma \left(\tau \right),
\quad
V_{logger} (\tau|x)= x^{\prime} \gamma \left(\tau^{\frac{1}{\lambda}} \right),
\label{Appmodel}
\end{equation}
where $x$ stacks the constant, appraisal value and volume of the auctioned tract.
In what follows, a median auction is an auction where the appraisal value and the volume are set to their median value. With the exception of Figure \ref{QFcoef}, all the figures and tables of this section and Appendix C are for a median auction.

\subsection{Specification analysis}
The fitted model (\ref{Appmodel}) combines a power asymmetry specification, i.e. $F_{logger} (v|x)=\left[F_{mill}(v|x)\right]^{\lambda}$ with a quantile regression for the parent distribution, which is identical here to the mill private value distribution. These two components are in fact quite different. Many options, such as adding interaction terms  or adopting a sieve approach as in \cn{BelloniChernozhukovChetverikovFernandez-Val2019}  can be used to improve the fit of the parent quantile regression. As $F_{logger} (v|x)=\text{Asy}\left[F_{mill}(v|x)|x\right]$ for the ``asymmetry'' function $\text{Asy}(\tau|x)=F_{logger} \left(V_{mill}(\tau|x)|x\right)$, the considered asymmetry power specification is quite restrictive and may fail to provide a good approximation for $\text{Asy}(\tau|x)$. It is therefore of interest  to develop a two-step analysis where the asymmetry specification is considered first, as this component of the model is the most likely to be misspecified.  The correct joint specification of the two components in (\ref{Appmodel}) is analyzed later.

\medskip

%\textsc{Belloni, A., V. Chernozhukov, D. Chetverikov \& I. Fernandes-V\'{a}l }(2019). Conditional quantile processes based on series or many regressors. \emph{Journal of Econometrics} \textbf{213}, 4--29.

\subsubsection{Asymmetry power specification \label{Asyspec}}

Let $P$ and $Q$ be the number of mills and loggers attending the auction. An implication of the asymmetry power specification already used for estimating $\lambda$ is
\[
\text{H}_0^{Asy}(p,q):
\quad
P\left(
\left. \text{The winner is a mill} \right| X, (P,Q)=(p,q)
\right)
=
\frac{p}{p+\lambda q}.
\]
Our power specification analysis is based on a test for  $\widehat{\text{H}}_0^{Asy}=\cup_{p,q}\text{H}_0^{Asy}(p,q)$, where the union is over the proportions $(p,q)$ with asymmetric auctions (i.e. $pq>0$, as the winner type distribution is degenerated otherwise), and with a number $L_{p,q}=\sum_{\ell=1}^{L} \mathbb{I} \left[(P_{\ell},Q_{\ell})=(p,q)\right]$ of auctions larger than $30$. A $t$ statistic for $\text{H}_0^{Asy}(p,q)$ is
\begin{align}
\widehat{\xi}_{p,q}
& =
\sqrt{L_{p,q}} \frac{\widehat{\omega}_{p,q} - \frac{p}{p+q\widehat{\lambda} }}{\widehat{\sigma}_{p,q}}
\label{Xipq}
\\
& 
\text{ where }
\widehat{\omega}_{p,q}
=
\frac{1}{L_{p,q}}
\sum_{\ell=1}^{L}
\mathbb{I}
\left(
\text{Auction $\ell$ winner is a mill and } (P_{\ell},Q_{\ell}) =(p,q)
\right)
\text{ and }
\nonumber \\
\widehat{\sigma}^2_{p,q}
& = 
\left(
\frac{
	\widehat{\omega}_{p,q} (1-\widehat{\omega}_{p,q})}{
	\sum_{s,t} 
	\frac{L_{s,t}}{L_{Asy}}
	\widehat{\omega}_{s,t} (1-\widehat{\omega}_{s,t})	
}
\frac{L_{p,q}}{L_{Asy}}-1
\right)^2
\widehat{\omega}_{p,q} (1-\widehat{\omega}_{p,q})
\nonumber \\
&
+ 
\left(
\frac{
	\widehat{\omega}_{p,q} (1-\widehat{\omega}_{p,q})}{
	\sum_{s,t} 
	\frac{L_{s,t}}{L_{Asy}}
	\widehat{\omega}_{s,t} (1-\widehat{\omega}_{s,t})	
}
\right)^2
\frac{L_{p,q}}{L_{Asy}}
\sum_{s,t:(s,t)\neq(p,q)} 
\frac{L_{s,t}}{L_{Asy}}
\widehat{\omega}_{s,t} (1-\widehat{\omega}_{s,t})	
\nonumber
\end{align}
where $L_{Asy}$ is the number of asymmetric auctions in the sample, i.e.  with  $P_{\ell} \neq 0$ or $Q_{\ell} \neq 0$. In (\ref{Xipq}), $\widehat{\omega}_{p,q}$ is the sample estimator of the probability that a mill wins in an auction with $p$ mills and $q$ loggers. The $t$-statistic $\widehat{\xi}_{p,q}$ is the studentized difference of $\widehat{\omega}_{p,q}$ to its model counterpart, which converges to a standard normal as shown in Proposition \ref{Xilim} in Appendix B under standard assumptions. 

Our asymmetry specification analysis relies on the maximum statistic over asymmetric auctions
\[
\max |\widehat{\xi}|
=
\max_{(p,q): L_{p,q}>30,pq\neq 0}
\left|\widehat{\xi}_{p,q} \right|,
\] 
which is used to test $\widehat{\text{H}}_0^{Asy}$. The $p$ value of $\max |\xi|$ is computed by the pairwise bootstrap as detailed in Appendix B. The result of this test is reported in Table \ref{Xitest}.
\begin{table}[!htbp]
	\centering
	\caption{ Asymmetry power specification}
	\begin{tabular}{lcc}
		\hline\hline 
		& Test statistic & $p$-value  \\
		\hline
		$\max |\widehat{\xi}|$ & 2.90 &  0.54 \\
		\hline
		\hline
		\label{Xitest}
	\end{tabular}
\end{table}
Appendix B also reports the pairwise bootstrap $\widehat{\xi}_{p,q}$ $p$-values, which are all reasonably high.

\subsubsection{Power and parent distribution specifications}

Our estimation strategy is built on the winning bid quantile function given the type proportion $(P,Q)$, the winner type, say $T$, and the auction characteristic $X$. More specifically, (\ref{Gamident}) shows that the conditional winning bid quantile is
\begin{equation}
W(\tau|X,P,Q,T,\gamma(\cdot),\lambda)= X^{\prime} \gamma \left( \Psi^{-1} (\tau|P,Q,T,\lambda)\right)
\label{Wapp}
\end{equation}
where, for $\lambda_T=1$ if the winner is a mill and $\lambda_T=\lambda$ if a logger wins and
$\Lambda_{P,Q}= P + \lambda Q$,
\[
\Psi (\tau|P,Q,T,\lambda)
=
\frac{\Lambda_{P,Q} \tau^{\Lambda_{P,Q}-\lambda_T}-\left(\Lambda_{P,Q}-\lambda_T\right)\tau^{\Lambda_{P,Q}}}{\lambda_T}.
\]
A recent literature considers quantile regression specification tests over a quantile interval $\mathcal{T}$, see \cn{EscancianoVelasco2010}, \cn{RotheWied2013}, \cn{EscancianoGoh2014} and the references therein. Their approach can be used to test whether the quantile regression (\ref{Wapp}) is correctly specified for each value of $(P,Q,T)$, building on a collection of statistics as done in the previous section for the winner type distribution.  We adopt instead a more aggregated approach which follows \cn{RotheWied2013} and 
avoids to estimate the conditional winning bid cdf given $(P,Q,T)$.
Let $G(w,x)=
\mathbb{E}
\left[
\mathbb{I}
\left(W \leq w \text{ and } X \leq x\right)
\right]
$\footnote{For a vector, $X\leq x$ means that $X_{j} \leq x_{j}$ for all $j$.}
be the joint cdf of the winning bid $W$ and $X$, which is estimated using the empirical cdf 
$
\widehat{G}(w,x) 
=
\frac{1}{L} \sum_{\ell=1}^{L}
\mathbb{I}
\left(
W_{\ell} \leq w, X_{\ell} \leq x
\right)
$.
The  null hypothesis is
\[
\text{H}_0: \text{There exists $\gamma(\cdot)$ and $\lambda$ such that } 
G(w,x|\gamma(\cdot),\lambda) =G(w,x)
\text{ for all $w,x$.} 
\] 
The null winning bid distribution is estimated using the winning bid quantile function (\ref{Wapp}) via
\begin{align}
\widehat{G}(w,x|\widehat{\gamma}(\cdot),\widehat{\lambda})
& =
\frac{1}{L}
\sum_{\ell=1}^{L}
\mathbb{I}
\left(X_{\ell} \leq x \right)
\int_{0}^{1}
\mathbb{I}
\left[
W\left(t|X_{\ell},P_{\ell},Q_{\ell},T_{\ell},\widehat{\gamma}(\cdot),\widehat{\lambda}\right)
\leq
w 
\right]
d t
.
\label{HatG0}
\end{align}
The Rothe and Wield (2013) statistic for $\text{H}_0$ is
\begin{equation}
RW=\sum_{\ell=1}^{L}
\left(
\widehat{G}(W_{\ell},X_{\ell}|\widehat{\gamma}(\cdot),\widehat{\lambda})
-
\widehat{G}(W_{\ell},X_{\ell})
\right)^2.
\label{RW}
\end{equation} 
	We have conducted in parallel a conditional testing procedure reported in Appendix B, which computes a statistic $RW$ as above for each type proportion observed in the sample, as done when analyzing the power specification. This was motivated by \cn{AradillasGandhiQuint2013}, who mentioned that auctions with $N=12$ bidders may have more bidders.\footnote{See their Footnote 26. This was also pointed to us by an anonymous Referee. These auctions represent slightly less than $8\%$ of the sample.} We therefore compute several versions of $RW$ depending on whether auctions with $N=12$ were used or not to estimate the parent distribution and $\lambda$, and in empirical cdf summations.
	Appendix B details how to compute (\ref{HatG0}) in practice and to apply the bootstrap procedure of \cn{RotheWied2013} to obtain the $p$-values of the next table.
	\begin{table}[!htbp]
		\centering
		\caption{ Asymmetry power and parent distribution specification}
		\begin{tabular}{l|cc|cc|cc}
			\hline\hline 
			& \multicolumn{2}{c}{$\widehat{\lambda},\widehat{\gamma}(\cdot)$: all sample} & \multicolumn{2}{|c}{$\widehat{\lambda},\widehat{\gamma}(\cdot)$: without $N=12$} & \multicolumn{2}{|c}{$\widehat{\lambda},\widehat{\gamma}(\cdot)$: without $N=12$} \\
			& \multicolumn{2}{c}{$\widehat{G}$: all sample} & \multicolumn{2}{|c}{$\widehat{G}$: all sample} & \multicolumn{2}{|c}{$\widehat{G}$: without $N=12$} \\
			& Test statistic & $p$-value  & Test statistic & $p$-value & Test statistic & $p$-value\\
			\hline
			RW & .098 &  0.043 & .120 & .094 &.085 & .170 \\
			\hline
			\hline
		\end{tabular}
		\label{RWtests}
\end{table}

	Table \ref{RWtests} shows that the Rothe and Wied (2013) test does not reject $\text{H}_0$ at the $1\%$ level but rejects at  $5\%$. Removing auctions with twelve bidders from the sample gives much higher p-values, which suggests that the proposed testing procedure supports the correct specification of the model. Further analysis reported in Appendix B shows that including or not auctions with twelve bidders gives a nearly identical estimation of $\lambda$ as well as the intercept and appraisal value coefficients, and only slightly increases the estimation of the volume slope. We therefore use the whole sample in the rest of the empirical analysis.

\subsection{Private value quantile functions}

We use a type fixed effect specification for the asymmetry parameter $\lambda_{i\ell}$, with $\lambda_{i\ell} = \lambda_{M}$ if bidder $i$ at auction $\ell$ is a mill and $\lambda_{i\ell} = \lambda_{L}$ if it is a logger. For identification, we normalize $\lambda_{M} = 1$. The first step estimation gives $\widehat{\lambda}_{L} = 0.6988$ with a 95\% confidence interval computed by pairwise bootstrap given by $\left[ 0.6516, 0.7554\right]$, which shows that loggers are indeed significantly weaker than mills. In particular, the logger winning probability  is $41.1\%$ when the types are in equal proportions, $70\%$ of the probability that a mill wins the ascending auction, which is $58.8\%$. 

\begin{figure}[!htbp]
\caption{Private Value Conditional Quantile Function of Loggers and Mills}
\subcaption*{Median Auction}
\begin{minipage}{0.8\linewidth}
\includegraphics[width=\linewidth]{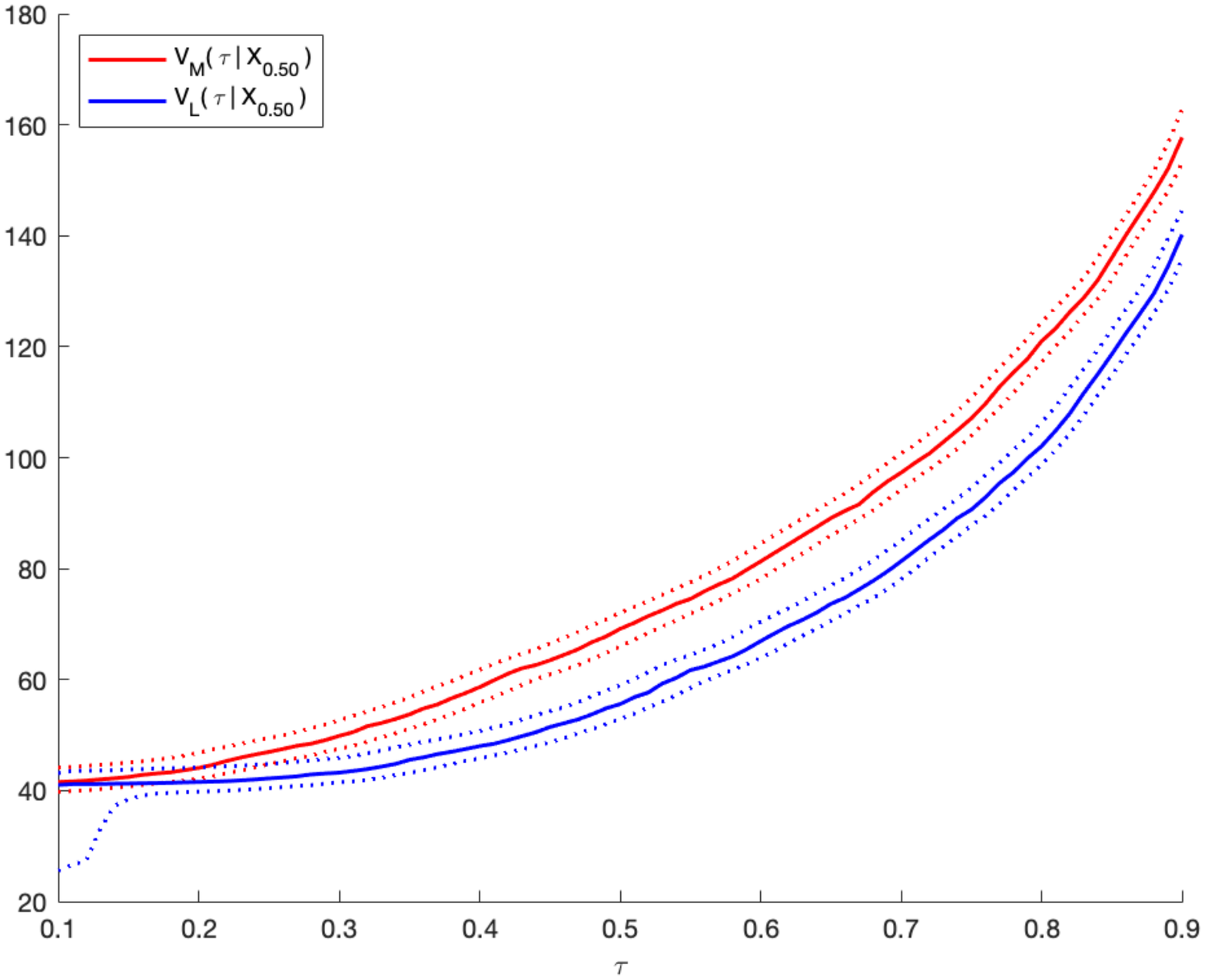}
\end{minipage} \label{QFtypes}
\begin{flushleft}  \scriptsize  The 95\% confidence intervals for the quantile regression estimates were computed by resampling with replacement the ($X_{\ell},W_{\ell}$)-pair. Mills in red and loggers in blue.\end{flushleft}
\end{figure}

This is confirmed by Figure \ref{QFtypes}, which gives the estimated private value quantile functions of mills (red) and loggers (blue) and their 95\% confidence bands computed via pairwise bootstrap method for a median auction. The private value conditional distribution of mills first-order stochastically dominates the one of loggers, especially in the upper part of the distribution.   

\FloatBarrier
\FloatBarrier

\begin{figure}[!htbp]
\caption{Private Value Parent Quantile Regression Coefficients}
\begin{minipage}{0.8\linewidth}
\includegraphics[width=\linewidth]{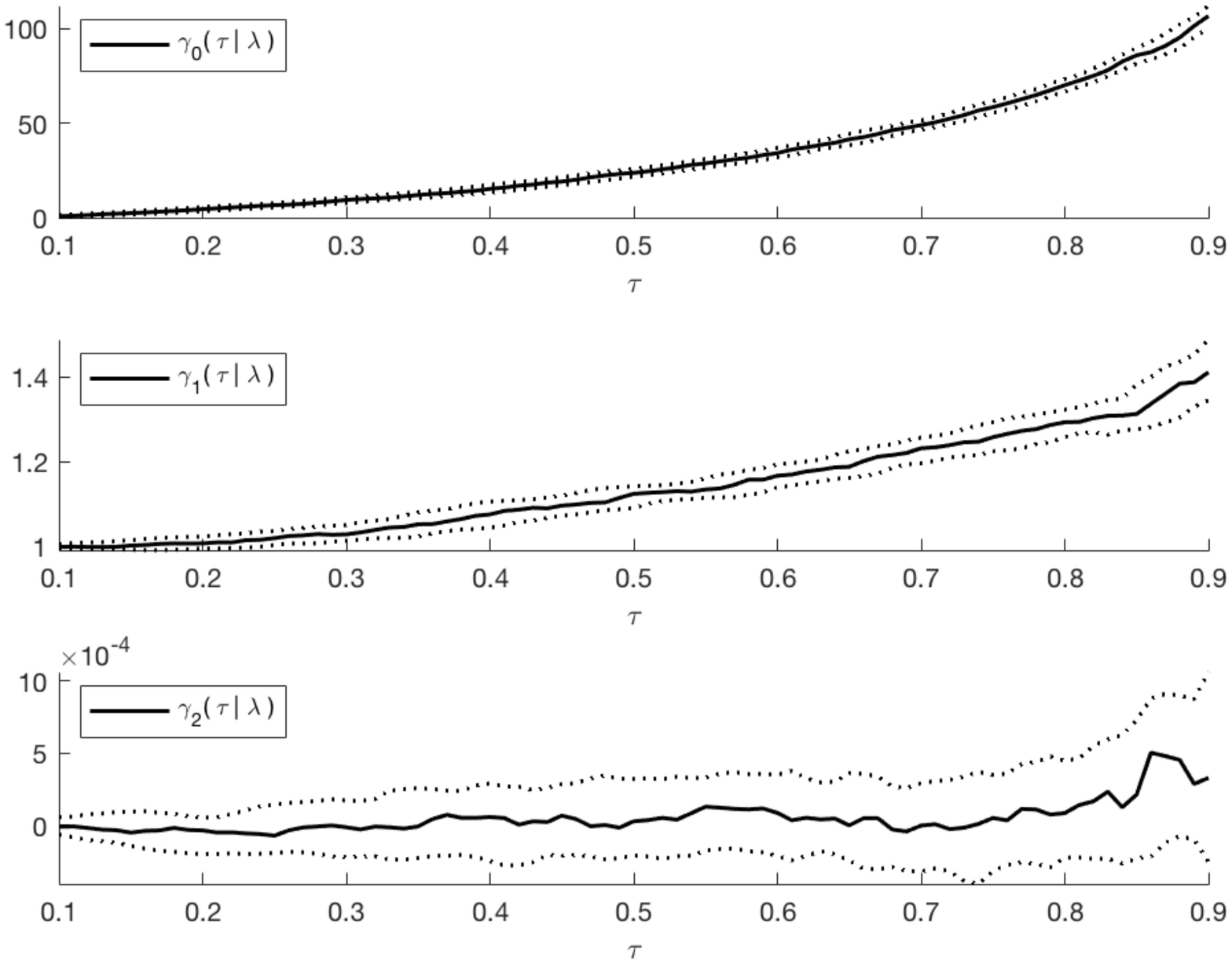}
\end{minipage} \label{QFcoef}
\begin{flushleft}  \scriptsize  The 95\% confidence intervals for the quantile regression estimates were computed by resampling with replacement the ($X_{\ell},W_{\ell}$)-pair. Top intercept, middle appraisal value and bottom volume estimated slope functions. \end{flushleft}
\end{figure}

The power specification of the private value quantile functions allows to highlight which variable generates asymmetry. Indeed, a constant slope function in the parent private value quantile regression means that the impact of the associated variable is identical for each type of bidders.  Figure \ref{QFcoef} gives the quantile regression coefficients of the private value parent distribution. The estimated volume slope function looks constant, and possibly not significant. As the power transformation will not make bidders to differ in terms of volume slope functions, this suggests that capacity constraint is not binding for both types. In contrast, the parent appraisal value slope function does not look constant, and this variable is much likely to generate differences across mills and loggers. Figure \ref{QFcoef}, therefore, suggests that asymmetry is driven by  qualitative (e.g. ability to improve on the appraisal value of the timber) and unobserved factors (captured by the intercept), instead of capacity constraints. Interestingly, coping for asymmetry gives appraisal value slope estimated functions that vary much less across quantile levels than in \cn{Gimenes2017}.

\FloatBarrier
\FloatBarrier

\subsection{Expected revenue and optimal reserve price \label{SER}}

We now investigate the effect of asymmetry on the seller's expected revenue and optimal reserve price. Given
that we recover all the primitives of the game, we can evaluate the seller expected revenue
as the proportion of types changes. This contrasts with \cn{CoeyLarsenSweeneyWaisman2017} who averages over the type proportion. In sections \ref{SER} and \ref{ABE}, the seller's outside option value $V_0$ appearing in Proposition \ref{ERRP} is set to $0$. Plotting  $r\in[0,1] \mapsto
\left(\widehat{V} (r|X),\widehat{\Pi}\left(  r|X\right)\right)$ gives a graph of the estimated seller's expected revenue achieved with a reserve price $R= \widehat{V} (r|X)$.

\begin{figure}[!htbp]
	\caption{Strategical Expected Revenue and Optimal Reserve Price}
	\begin{minipage}{1.0\linewidth}
		\includegraphics[width=\linewidth]{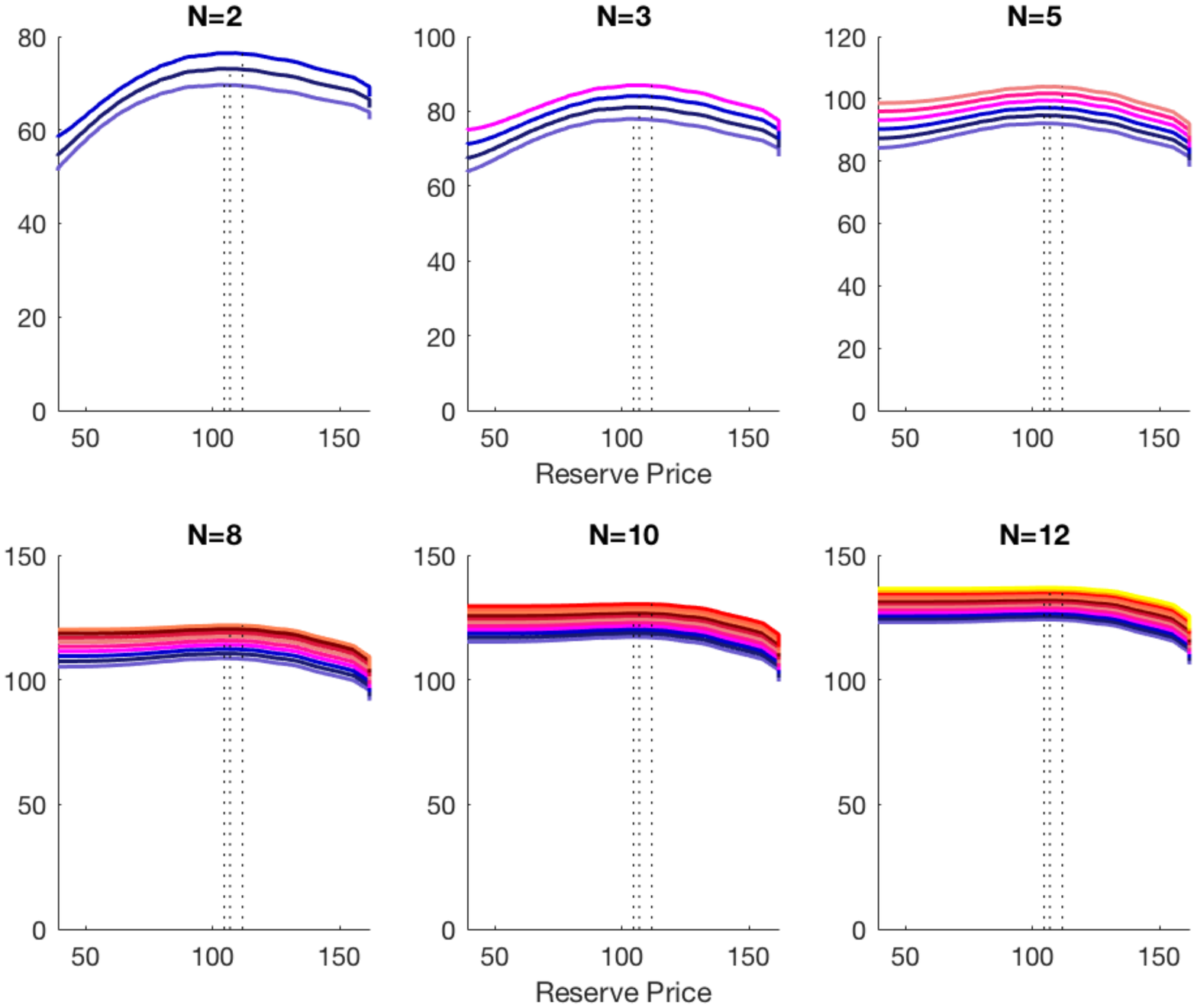}
	\end{minipage} \label{ER_ORP}
\end{figure}

Figure \ref{ER_ORP} shows estimates of the expected revenue as a function of the reserve price for each $N$ and type proportion. The dotted vertical lines give the optimal reserve price for each proportion of types.	
As the colors of the curves become warmer (from blue to red and yellow), loggers are replaced by mills and the revenue level increases in a parallel way.  The expected revenue functions have clear maximas for small numbers of bidders (typically $N=2$ or $N=3$), contrasting with the estimation obtained with symmetric bidders in
\cn{GimenesGuerre2019}.
For larger $N$, the expected revenues look flat in their central part, a fact that cannot be seen from the estimation set strategy of \cn{CoeyLarsenSweeneyWaisman2017}.

As a consequence, implementing an optimal reserve price is mostly useful when the probability of observing a small number of bidders is high. The optimal reserve prices shown in Figure \ref{ER_ORP} and detailed in the Appendix Table \ref{count_prop_ORP} depend upon $N$ and type proportion, but exhibit a moderate $7\%$ variation, staying in the interval $[104.7,111.9]$ and slightly increasing with the number of mills. As the expected revenues are flat around their maxima, using a reserve price in the range $[104.7,111.9]$ gives an expected revenue close to its maxima. This includes the optimal reserve price $107.9\$$ estimated from a symmetric specification, as in \cn{Gimenes2017}, given in \ref{count_prop_ORP}. As the expected revenue with no reserve price is mostly below $100\$$ when $N \leq 5$, as seen from Table \ref{ERchanges_nonstrat} below, using such a reserve price may mean not selling the auctioned lot if a small number of bidders participates.\footnote{To see this, observe that the probability of selling is the probability that the maximum private value $V_{(N)}$ is above the reserve price $R$. The Markov inequality gives the bound $\mathbb{E}[V_{(N)}]/R$ for the latter. A proxy for $\mathbb{E}[V_{(N)}]/R$ is the non strategical revenue $\Pi (0)$ when the seller value is $0$, suggesting to use the bound $\Pi(0)/R$ for the probability of selling.}   
\FloatBarrier
\FloatBarrier

\subsection{Type variation and additional bidder  effects \label{ABE}}

\begin{table}[htbp]
	\centering
	\caption{Non Strategical Expected Revenue}
	\begin{tabular}{lcccc}
		\hline\hline
		&             &                &                            & One logger replaced by one mill \\
		& Min ER & Max ER  & Max $\%\Delta$ & [ Min \%, Max \% ] \\
		\hline
		$N=2$	&48.02&	57.65& 19.61\% & [ 8.84\%, 9.89\% ] \\
		&	\small{[46.37, 50.42]} &	\small{[56.20, 59.70]} &  &\\	
		$N=3$ &	63.59 &75.04	 & 18.01\% & [ 5.42\%, 5.83\% ] \\
		&	\small{[61.61, 66.03} &	\small{[73.30, 77.47]} &  & \\	
		$N=5$ &	84.25 &	98.64 &	 17.08\% & [ 2.78\%, 3.63\% ]\\
		&	\small{[81.82, 87.14]} &	\small{[96.51, 101.45]} &   &\\	
		$N=8$ &	105.28 &	120.17 & 14.14\% & [ 1.35\%, 2.02\% ]\\
		&	\small{[102.57, 108.42]} &	\small{[117.68, 123.34]} &   &\\	
		$N=10$ &	115.22 &	129.55 & 12.44\% & [ 0.93\%, 1.48\% ]\\
		&	\small{[112.45, 118.58]} &	\small{[126.59, 132.95]} &  &\\	
		$N=12$ &	123.05 &	136.52 & 10.95\% & [ 0.66\%, 1.12\% ]\\
		&	\small{[120.23, 126.19]} &	\small{[133.61, 139.96]} &  &\\	
		\hline
		\hline
		\label{ERchanges_nonstrat}
	\end{tabular}
	\begin{flushleft}  \scriptsize  The 95\% confidence intervals for the quantile regression estimates were computed by resampling with replacement the ($X_{\ell},W_{\ell}$)-pair. \end{flushleft}
\end{table}

\begin{table}[htbp]
	\centering
	\caption{Strategical Expected Revenue}
	\begin{tabular}{lcccc}
		\hline\hline
		&             &                &                            & One logger replaced by one mill \\
		& Min ER & Max ER  & Max $\%\Delta$ & [ Min \%, Max \% ] \\
		\hline
		$N=2$	&69.65&	76.44 & 9.75\% & [ 4.57\%, 4.95\% ] \\
		&	\small{[68.64, 70.92]} &	\small{[75.45, 77.77]}&  &\\	
		$N=3$	&	77.97 &86.95 & 11.52\% & [ 3.44\%, 3.98\% ]\\
		&		\small{[76.60, 79.67]} &	\small{[85.62, 88.73]}& &\\	
		$N=5$	&	92.16 &	103.9  & 12.74\% & [ 2.14\%, 2.75\% ]\\
		&		\small{[90.27, 94.45]} &	\small{[102.06, 106.30]} &  &\\	
		$N=8$	&	108.63 &	121.86  & 12.18\% &[ 1.20\%, 1.73\% ] \\
		&		\small{[106.32, 111.36]} &	\small{[119.54, 124.84]} & &\\	
		$N=10$	&	117.16 &	130.37  & 11.28\% & [ 0.86\%, 1.33\% ] \\
		&		\small{[114.70, 119.99]} &	\small{[127.76, 133.61]} &  &\\	
		$N=12$	&	124.2 &	136.92  & 10.24\% & [ 0.63\%, 1.03\% ] \\
		&		\small{[121.63, 127.20]} &	\small{[134.11, 140.35]} &  &\\	 
		\hline
		\hline
		\label{ERchanges_strat}
	\end{tabular}
	\begin{flushleft}  \scriptsize  The 95\% confidence intervals for the quantile regression estimates were computed by resampling with replacement the ($X_{\ell},W_{\ell}$)-pair. \end{flushleft}
\end{table}

In this section, we study the effects of changes in the bidder's type proportion and additional bidders on the expected revenue. For that, we set the largest $N$ to its maximal observed value 12, see tables \ref{count_prop_ER_nonstr} and \ref{count_prop_ER} in Appendix C.
The $95\%$ boostrapped confidence intervals for the expected revenue given in these tables have a length ranging from $2\$$ to $6\$$, corresponding to revenues varying between $48\$$ and $137\$$
\footnote{The bootstrap $95\%$ confidence intervals for the optimal reserve price 
have a larger length, between $12\$$ and $14\$$ for an optimal reserve price between $104\$$ and $112\$$.
As a matter of comparison, \cn{CoeyLarsenSweeneyWaisman2017}'s set identified confidence bounds for the seller revenue and optimal reserve price look huge, but they also allow for affiliated values.}.
The bootstrapped $95\%$ confidence intervals of the strategical seller expected revenue, achieved using an optimal reserve price, and the non strategical one, obtained with a non binding reserve price, does not overlap up to $N=6$.	
Similarly, the revenue gain achieved when an additional bidder of any type enters looks significant, at least for auctions with up to $7$ initial bidders for additional logger and, for mill, up to some auctions with $N=10$. Setting the largest $N$ to 12 is therefore expected to capture all the statistically significant policy effects delivered by the sample. We now focus on each of these effects.

\paragraph{Revenue and types.}
Point estimation of bidders' private value distributions permits investigation of changes in the number of bidders of a given type. 
Tables \ref{ERchanges_nonstrat}, \ref{ERchanges_strat} and \ref{BK} give a summary of all universe of changes, see also Tables \ref{count_prop_ER_nonstr} and \ref{count_prop_ER}  in Appendix C.

Table \ref{ERchanges_nonstrat} considers a non strategical expected revenue, which means that reserve price is non binding, whereas Table \ref{ERchanges_strat} focuses on the optimal revenue\footnote{As suggested in \cn{CoeyLarsenSweeney2019}, a comparison between a strategical and non strategical expected revenue can be fruitful to the seller due to the costs that a policy of setting an optimal reserve price may impose in practice. Recent works have highlighted the asymmetric effects on seller's revenue due to mistakes in choosing reserve prices (see e.g. \cn{Kim2013}, \cn{OstrovskySchwarz2016}, \cn{CoeyLarsenSweeney2019} and \cn{Gimenes2017})}. All the results are obtained for a given $N$. The second and third columns of both tables give the minimum and maximum values of the seller expected revenue across type proportions. The minimum and maximum values of the revenue in both cases are obtained when only loggers and only mills are participating, respectively. The percentage change in revenue when changing all loggers into mills is given in the fourth column and is an additional measure of asymmetry. It is, on average, 15.4\% in the non strategical case and 11.3\% in Table \ref{ERchanges_strat}. These order of magnitude are similar to the one found in \cn{RobertsSweeting2016} who employ a parametric specification.\footnote{These authors also allow for entry decision but their estimate ``indicate a moderate effect of selection''.} The fifth column gives the maximum and minimum percentage changes obtained when replacing one logger by one mill. All these results suggest that the seller should either incentivize mills participation or subsidize higher loggers bid as studied in \cn{FlambardPerrigne2006}, \cn{Marion2007} or \cn{KrasnokutskayaSeim2011} for the latter.

\begin{table}[!htbp]
	\centering
	\caption{Violations of Bulow and Klemperer (1996), $N=2,3,4$}
	\begin{tabular}{cclccccc}
		\hline\hline
		$N$ & (Logger,Mill) && Non strat. ER & Strat. ER & Additional Logger & Additional Mill \\
		\hline
		$N=2$ &  (2,0) & \vline & 48.02 & 69.65 & $\underline{63.59}^*$ &  $\underline{67.30}$ \\
		      &  (1,1) & \vline & 52.46 & 73.10 & $\underline{67.30}^*$ & $\underline{71.18}$ \\
		      &  (0,2) & \vline & 57.65 & 76.44 & $\underline{71.18}^*$ & $\underline{75.04}$ \\
		\hline
		$N=3$ & (3,0) & \vline & 63.59 & 77.97 & $\underline{74.82}$ & $78.23$ \\
		      & (2,1) & \vline & 67.30 & 81.07 & $\underline{78.23}$ & $81.63$ \\
		      & (1,2) & \vline & 71.18 & 84.06 & $\underline{81.63}$ & $84.96$  \\
		      & (0,3) & \vline & 75.04 & 86.95 & $\underline{84.96}$ & $88.17$  \\
		\hline
		$N=4$ & (4,0) & \vline & 74.82 & 85.44 & $\underline{84.25}$ & $87.31$ \\
		      & (3,1) & \vline & 78.23 & 88.24 & $\underline{87.31}$ & $90.30$ \\
		      & (2,2) & \vline & 81.63 & 90.93 & $\underline{90.30}$ & $93.19$ \\
		      & (1,3) & \vline & 84.96 & 93.53 & $\underline{93.19}$ & $95.97$ \\
		      & (0,4) & \vline & 88.17 & 96.03 & $\underline{95.97}$ & $98.64$ \\
		\hline
		\hline
		\label{BK}
	\end{tabular}
\begin{flushleft}  \scriptsize
	An underlined revenue indicates a violation of \cn{BulowKlemperer1996}, ie the considered non strategical revenue obtained by adding a bidder of a given type is below the strategical one. A ``*'' indicates that the $95\%$ bootstrapped confidence interval of the strategical revenue and the non strategical one with an additional bidder of the considered type do not overlap.
\end{flushleft}
\end{table}

\paragraph{Revenue and additional bidders.}

An important result by \cn{BulowKlemperer1996} states that the seller expected revenue achieved in an ascending auction with no reserve price but an additional bidder is higher than the one of any allocation mechanism, which includes the case of an ascending auction with an optimal reserve price, under symmetry and a downward sloping marginal revenue condition.\footnote{See \cn{CoeyLarsenSweeney2019} for a recent econometric application to entry exogeneity.} 
Table \ref{BK} reports several violations of \cn{BulowKlemperer1996} arising in our asymmetric framework.
The ``Strat. ER'' column of Table \ref{BK} indicates the estimated optimal expected revenue achieved with $N=2,3$ and $4$ bidders, with number of loggers or mills as indicated in the second column. The last two columns give the estimated non strategical expected revenue obtained when adding a logger or a mill.

Table \ref{BK} shows that using an optimal reserve price is always more profitable than adding a weak logger bidder. Adding a mill bidder is also less profitable than using the optimal auction but only when $N=2$ and in a much less significant way than adding a logger. Table \ref{BK} shows that the difference of revenue using the optimal auction and adding a logger decreases with $N$, in average across type proportion. By contrast the revenue difference using the optimal auction and adding a mill increase with $N$.\footnote{Tables \ref{count_prop_ER_nonstr} and \ref{count_prop_ER} in Appendix C also report the revenues obtained for an estimation of a symmetric private value model as in \cn{Gimenes2017}. Interestingly violations of \cn{BulowKlemperer1996} occur for $N=2,3$ but not for larger $N$.}
The systematic violations of \cn{BulowKlemperer1996} when adding a logger suggests that the logger private value distribution does not satisfy the downward sloping marginal revenue condition.\footnote{The downwards sloping marginal revenue condition of \cn{BulowKlemperer1996} requires that 
	\[
	-
	\frac{d}{dt}
	\left[
	V_i (t) (1-t)\right]
	=
	V
	\left(t^{1/\lambda_i}\right)
	-
	(1-t)
	\frac{t^{1/\lambda_i-1}}{\lambda_i}
	V^{(1)}
	\left(t^{1/\lambda_i}\right)
	\]
	increases with $t$. If $V^{(1)} (0)>0$ and  $1/2<\lambda_i<1$, the leading term when $t$ goes to 0 of the derivative of this function is
	$
	-(1/\lambda_i-1) \frac{t^{1/\lambda_i-2}}{\lambda_i} V^{(1)} (0)
	$
	which is negative, so that the considered condition is not compatible with our estimation of $\lambda_L$.	
	}
When $N \geq 4$, using the optimal reserve price is less profitable than participation of an additional bidder of any type, up to few minor exceptions. However, the differences of expected revenue between an optimal reserve price and an additional bidder are at best in the range of $3\$$, which is close to the half length of the boostrapped $95\%$ confidence interval for the strategical and non strategical seller's expected revenues.

\section{Conclusion}

The paper considers a semiparametric specification for asymmetric private value distribution under the independent private value distribution setup. The bidders share a common parent distribution, which is generated by a quantile regression model. Asymmetry is driven by powers applied to the parent distribution. These powers can depend upon individual and/or group fixed effects, bidder and/or auction specific variables. The specification can be estimated by a two stage procedure from the winning bid and winner's identity. This quantile regression specification is not affected by the curse of dimensionality and can cope with data-rich environment. Unlike common parametric specifications, it is expected to be less affected by misspecification due to its nonparametric nature. Usual parametric rates nevertheless apply and estimation techniques remain standard. The parametric power component of the model allows for a simple evaluation of bidder's asymmetry and of its economic implications.

A timber auction application has been used to illustrate the implication of asymmetry. The proposed specification tests do not reject the model.
The estimated asymmetry parameter means that weaker bidders have $30\%$ less chances to win the auction than stronger ones.
The quantile regression specification allows to detect the variables that affect the bidders in a symmetric way, here volume, suggesting that bidders face similar capacity constraints, and the other variables that represent characteristics of asymmetry. The shape of the expected revenue varies a lot with the number $N$ of bidders, being mostly flat for $N >5$, with an optimal revenue close to the one achieved in the absence of a reserve price. For small $N$, the choice of a reserve price does matter, but the estimated optimal one does not vary too much with $N$ and type proportion. The effect of asymmetry is mild here, and using the one estimated from a misspecified symmetric model should protect the seller against revenue loss occurring for small $N$.
On the other hand, and as expected, the proportion of small bidders may importantly affect the seller expected revenue. This suggests that the seller can benefit from preference policies which would strengthen the weak bidders. A striking finding is that, in small auctions with less than four bidders, increasing participation, as recommended by \cn{BulowKlemperer1996} in a symmetric environment, may give a smaller revenue than using an optimal reserve price, due to the presence of weak bidders. As a consequence, the choice of a proper reserve price may be a more important tool under asymmetry than when the bidders are symmetric.

\bibliography{bibfile}

\renewcommand{\theequation}{A.\arabic{equation}}
\renewcommand{\thesubsection}{A.\arabic{subsection}}

\section*{Appendix A - Proof section} \label{proof sect}

\subsection{Proof of Lemma \ref{Ident1}. }
Let
\[
G\left(  w,i|X,Z,N\right)  =\mathbb{P}\left(  \left.  W\leq w\text{ and }i\text{
	wins the auction}\right\vert X,Z,N\right)
\]
be the joint distribution of winning bids and winner's identity. Due to
private value independence, it holds, as shown in \cn{BendstrupPaarsch2006},
\begin{align*}
&  G\left(  w,i|X,Z,N\right)  =\mathbb{P}\left(  \left.  \max_{1\leq j\neq i\leq
	N}V_{j}\leq w\ \text{and }\max_{1\leq j\neq i\leq N}V_{j}\leq V_{i}\right\vert
X,Z,N\right) \\
&  \quad=\mathbb{P}\left(  \left.  \max_{1\leq j\neq i\leq N}V_{j}\leq
\min\left(  w,V_{i}\right)  \right\vert X,Z,N\right) \\
&  \quad=\int_{0}^{w}\left\{
%TCIMACRO{\dprod \limits_{1\leq j\neq i\leq N}}%
%BeginExpansion
{\displaystyle\prod\limits_{1\leq j\neq i\leq N}}
%EndExpansion
F_{j}\left(  u|X,Z_{j}\right)  \right\}  dF_{i}\left(  u|X,Z_{i}\right)
+\left\{
%TCIMACRO{\dprod \limits_{1\leq j\neq i\leq N}}%
%BeginExpansion
{\displaystyle\prod\limits_{1\leq j\neq i\leq N}}
%EndExpansion
F_{j}\left(  w|X,Z_{j}\right)  \right\}  \left(  1-F_{i}\left(  w|X,Z_{i}%
\right)  \right) \\
&  \quad=\int_{0}^{w}\left(  1-F_{i}\left(  u|X,Z_{i}\right)  \right)
d\left\{
%TCIMACRO{\dprod \limits_{1\leq j\neq i\leq N}}%
%BeginExpansion
{\displaystyle\prod\limits_{1\leq j\neq i\leq N}}
%EndExpansion
F_{j}\left(  u|X,Z_{j}\right)  \right\}
\end{align*}
where the last line is obtained by integration by parts. Then, Assumption \ref{Fi}
gives%
\begin{align*}
G\left(  w,i|X,Z,N\right)   &  =\int_{0}^{w}\left(  1-\left[  F\left(
u|X\right)  \right]  ^{\lambda_{i}}\right)  d\left[  F\left(  u|X\right)
\right]  ^{\sum_{1\leq j\neq i\leq N}\lambda_{j}}\nonumber\\
&  =
\left[  F\left(  w|X\right)  \right]  ^{\sum_{1\leq j\neq i\leq N}%
	\lambda_{j}}
-
\frac{\sum_{1\leq j\neq i\leq N}\lambda_{j}}{\sum_{j=1}%
	^{N}\lambda_{j}}\left[  F\left(  w|X\right)  \right]  ^{\sum_{j=1}^{N}%
	\lambda_{j}}
.
\end{align*}
It follows
\begin{align*}
&  \mathbb{P}\left(  \left.  \text{Bidder }i\text{ wins the auction}%
\right\vert X,Z,N\right)  =G\left(  +\infty,i|X,Z,N\right) \\
&  \quad=1-\frac{\sum_{1\leq j\neq i\leq N}\lambda_{j}}{\sum_{j=1}^{N}%
	\lambda_{j}}=\frac{\lambda_{i}}{\sum_{j=1}^{N}\lambda_{j}}
\end{align*}
and
\begin{eqnarray*}
G(w|X,Z,N,i)
& = &
\frac{G(w,i|X,Z,N)}{G(+\infty,i|X,Z,N)}
\\
& = &
\frac{
\left(
\sum_{j=1}^{N}
\lambda_{j}
\right)
\left[  
F\left(  w|X\right)  
\right]^{
\sum_{1\leq j\neq i\leq N}
\lambda_{j}
}
-
\left(
\sum_{1\leq j\neq i\leq N}
\lambda_{j}
\right)
\left[  
F\left(  w|X\right)  \right]^{\sum_{j=1}^{N}\lambda_{j} }
}{
\lambda_i}
\\
& = &
\Psi_{i} \left[ F(w|X)\right]
\end{eqnarray*}
where $\Psi_i (\cdot)$ is as in (\ref{Psii}).
This ends the proof of the Lemma.
\hfill $\Box$

\subsection{Proof of Proposition \ref{Ident2}}

Assumption \ref{Lambda} yields that the function $\Psi_{i}\left(
\tau;Z,\alpha,\beta\right)  =\Psi_{i}\left(  \tau\right)  $ is well-defined as
$\lambda\left(  Z_{i};\alpha_{i},\beta\right)  =\lambda_{i}>0$. Set
$\Lambda_{N}=\Lambda_{N}\left(  Z;\alpha,\beta\right)  $, $\Lambda
_{N|i}=\Lambda_{N|i}\left(  Z;\alpha,\beta\right)  =\Lambda_{N}-\lambda_{i}$
so that%
\begin{align*}
\Psi_{i}\left(  \tau\right)   &  =\frac{\Lambda_{N}\tau^{\Lambda_{N|i}%
}-\Lambda_{N|i}\tau^{\Lambda_{N}}}{\lambda_{i}},\\
\frac{\partial\Psi_{i}\left(  \tau\right)  }{\partial\tau}  &  =\frac
{\Lambda_{N}\Lambda_{N|i}\tau^{\Lambda_{N|i}-1}}{\lambda_{i}}\left(
1-\tau^{\lambda_{i}}\right)  .
\end{align*}
Hence $\Psi_{i}\left(  \cdot\right)  $ is continuous and strictly increasing.
(\ref{Gamident}) and Assumption \ref{X} then yield%
\[
\gamma\left(  \tau\right)  =\mathbb{E}^{-1}\left[  XX^{\prime}\right]
\times\mathbb{E}\left[  XW\left(  \Psi_{i}\left(  \tau\right)  |X,Z,i\right)
\right]
\]
for all $\tau$ in $\left[  0,1\right]  $.$\hfill\square$

\subsection{Proof of Proposition \ref{ERRP}}

Ignore, for the sake of brevity, the conditioning variables. Under assumption
\ref{R}, the seller possible payoffs are
\[
\pi\left(  r\right)  =\left\{
\begin{array}
[c]{cl}%
V_{0} & \text{if }V_{N:N}<R\\
R & \text{if }V_{N-1:N}<R\leq V_{N:N},\\
V_{N-1:N} & \text{if }R\leq V_{N-1:N},
\end{array}
\right.
\]
where $V_{i:N}$ is the $i$th-lowest order statistics of private values, i.e.
$V_{N:N}$ is the first highest order statistic and $V_{N-1:N}$ the second. Recall that $r=F\left(  R\right)  $, or equivalently $R=V\left(
r\right)  $. The next three points evaluate the contribution of each of the three events above to the seller revenue.

\begin{enumerate}
\item[1.] $\mathbb{P}\left(  V_{N:N}<R\right)  =\mathbb{P}\left(
V_{i}<R,\forall i=1,\ldots,N\right)  =\Pi_{i=1}^{N}F_{i}\left(  R\right)
=\Pi_{i=1}^{N}\left[  F\left(  V\left(  r\right)  \right)  \right]
^{\lambda_{i}}=r^{\Lambda_{N}}$. It follows that the probability of selling is $1-r^{\Lambda_{N}}$, hence, Proposition \ref{ERRP}-(i) is proven. The
contribution of this event to the seller revenue is $\pi_{1}(r)=V_{0}%
r^{\Lambda_{N}}$;

\item[2.] $\mathbb{P}\left(  V_{N-1:N}<R\leq V_{N:N}\right)  =\sum_{i=1}%
^{N}\Pi_{j\neq i}F_{j}\left(  R\right)  \left(  1-F_{i}\left(  R\right)
\right)  =\sum_{i=1}^{N}r^{\Lambda_{N|i}}(1-r^{\lambda_{i}})$. The
contribution of this second event to the seller revenue is $\pi_{2}%
(r)=V(r)\sum_{i=1}^{N}r^{\Lambda_{N|i}}(1-r^{\lambda_{i}})$;

\item[3.] Let $F_{N-1:N}(v)$ denote the c.d.f. of the second-highest order
statistic $V_{N-1:N}$, which is
\[
F_{N-1:N}(v)=\prod_{i=1}^{N}F_{i}(v)+\sum_{i=1}^{N}\prod_{j\neq i}%
F_{j}(v)(1-F_{i}(v)).
\]
Under Assumption \ref{Fi}
\begin{align*}
F_{N-1:N}(v)  &  =[F(v)]^{\Lambda_{N}}+\sum_{i=1}^{N}\left[
(1-(F(v))^{\lambda_{i}}).(F(v))^{\Lambda_{N|i}}\right] \\
&  =[F(v)]^{\Lambda_{N}}+\sum_{i=1}^{N}[(F(v))^{\Lambda_{N|i}}-(F(v))^{\Lambda
_{N}}]\\
&  =(1-N)(F(v))^{\Lambda_{N}}+\sum_{i=1}^{N}(F(v))^{\Lambda_{N|i}}.
\end{align*}
The change of variable $v=V(t)$ with $R=V(r)$ then gives that the contribution of the third
event to $\pi\left(  r\right)  $ is
\begin{align*}
\pi_{3}\left(  r\right)   &  =\int_{R}^{V(1)}vdF_{N-1:N}(v)=\int_{r}^{1}V\left(
t\right)  d\left[  (1-N)t^{\Lambda_{N}}+\sum_{i=1}^{N}t^{\Lambda_{N|i}}\right]
\\
&  =\int_{r}^{1}V\left(  t\right)  \left\{  (1-N)\Lambda_{N}t^{\Lambda_{N}%
-1}+\sum_{i=1}^{N}\Lambda_{N|i}t^{\Lambda_{N|i}-1}\right\}  dt.
\end{align*}

\end{enumerate}

As $\Pi\left(  r\right)  =\pi_{1}\left(  r\right)  +\pi_{2}\left(  r\right)
+\pi_{3}\left(  r\right)  $, Proposition \ref{ERRP}-(ii) is proved. It also
follows that%

\begin{align*}
\frac{\partial\Pi\left(  r\right)  }{\partial r}  &  =V_{0}\Lambda
_{N}r^{\Lambda_{N}-1}+V^{\left(  1\right)  }(r)\sum_{i=1}^{N}r^{\Lambda_{N|i}%
}(1-r^{\lambda_{i}})+R\sum_{i=1}^{N}\left(  \Lambda_{N|i}r^{\Lambda_{N|i}-1
}-\Lambda_{N}r^{\Lambda_{N}-1}\right) \\
&  -R\left\{  \Lambda_{N}r^{\Lambda_{N}-1}+\sum_{i=1}^{N}\left(  \Lambda
_{N|i}r^{\Lambda_{N|i}-1}-\Lambda_{N}r^{\Lambda_{N}-1}\right)  \right\} \\
&  =\left(  V_{0}-R\right)  \Lambda_{N}r^{\Lambda_{N}-1}+V^{\left(  1\right)
}(r)\sum_{i=1}^{N}r^{\Lambda_{N|i}}(1-r^{\lambda_{i}})\\
&  =\Lambda_{N}r^{\Lambda_{N}-1}\left\{  V_{0}-R-\frac{V^{\left(  1\right)
}(r)r}{\Lambda_{N}}\sum_{i=1}^{N}(1-r^{-\lambda_{i}})\right\}  .
\end{align*}
Note that the optimal $r_{\ast}$ must belong to the open set $\left(
0,1\right)  $. Hence the FOC\ $\frac{\partial\Pi\left(  r_{\ast}\right)
}{\partial r}=0$ gives that Proposition \ref{ERRP}-(iii) holds.$\hfill\square$

\subsection{Proof of Theorem \ref{AN}}

By Theorems 2.5 and 3.3, the proof of Theorem 3.2 in \cn{NeweyMcFadden1994}, it holds under Assumptions \ref{Lambda}, \ref{Ind} and \ref{FSident}%
\begin{equation}
\sqrt{L}\left(  \widehat{\theta}-\theta\right)  =\widehat{\Sigma
}+o_{\mathbb{P}}\left(  1\right)  ,\quad\widehat{\Sigma}=\mathcal{I}\left(
\theta\right)  ^{-1}\frac{1}{\sqrt{L}}\sum_{\ell=1}^{L}\frac{P^{\theta
}\left(  I_{\ell}^{\ast}|Z_{\ell},N_{\ell},\theta\right)  }{P\left(
I_{\ell}^{\ast}|Z_{\ell},N_{\ell},\theta\right)  }+o_{\mathbb{P}}\left(
1\right)  . \label{FS}%
\end{equation}
For $\widehat{\gamma}\left(  \tau\right)  $, define%
\[
\widehat{Q}\left(  \gamma;\vartheta\right)  =\frac{1}{L}\sum_{\ell=1}^{L}%
\rho_{\Psi_{I_{\ell}^{\ast}}\left(  \tau;Z_{\ell},N_{\ell},\vartheta\right)
}\left(  W_{\ell}-X_{\ell}^{\prime}\gamma\right)
\]
which is such that $\widehat{\gamma}\left(  \tau\right)  =\arg\min_{\gamma
}\widehat{Q}\left(  \gamma;\widehat{\theta}\right)  $ and $\widetilde{\gamma
}\left(  \tau\right)  =\arg\min_{\gamma}\widehat{Q}\left(  \gamma
;\theta\right)  $. The proof makes use of the following partial derivatives%

\begin{align*}
\widehat{Q}_{\vartheta}  &  =\left.  \frac{\partial\widehat{Q}\left(
\gamma;\vartheta\right)  }{\partial\vartheta}\right\vert _{\left(
\gamma,\vartheta\right)  =\left(  \gamma\left(  \tau\right)  ,\theta\right)
}=\frac{1}{L}\sum_{\ell=1}^{L}\left(  W_{\ell}-X_{\ell}^{\prime}\gamma\left(
\tau\right)  \right)  \Psi_{I_{\ell}^{\ast}}^{\theta}\left(  \tau;Z_{\ell
},N_{\ell},\theta\right)  ,\quad Q_{\vartheta}=\mathbb{E}\left[
\widehat{Q}_{\vartheta}\right]  ,\\
\widehat{Q}_{\vartheta\vartheta}  &  =\frac{1}{L}\sum_{\ell=1}^{L}\left(
W_{\ell}-X_{\ell}^{\prime}\gamma\left(  \tau\right)  \right)  \Psi_{I_{\ell
}^{\ast}}^{\theta\theta}\left(  \tau;Z_{\ell},N_{\ell},\theta\right)  ,\quad
Q_{\vartheta\vartheta}=\mathbb{E}\left[  \widehat{Q}_{\vartheta\vartheta
}\right] \\
\widehat{Q}_{\vartheta\gamma}  &  =-\frac{1}{L}\sum_{\ell=1}^{L}\Psi_{I_{\ell
}^{\ast}}^{\theta}\left(  \tau;Z_{\ell},N_{\ell},\theta\right)  X_{\ell
}^{\prime},\quad D\left(  \tau\right)  =\mathbb{E}\left[  \widehat{Q}%
_{\vartheta\gamma}\right]  .
\end{align*}
Let $\widehat{S}/\sqrt{L}=\widehat{S}\left(  \tau\right)/\sqrt{L}  $ be the
$\gamma$-derivative of $\widehat{Q}\left(  \gamma;\theta\right)  $ taken at
$\gamma\left(  \tau\right)  $%
\[
\widehat{S}=\frac{1}{\sqrt{L}}\sum_{\ell=1}^{L}X_{\ell}\left[  \mathbb{I}%
\left(  W_{\ell}\leq X_{\ell}^{\prime}\gamma\left(  \tau\right)  \right)
-\Psi_{I_{\ell}^{\ast}}\left(  \tau;Z_{\ell},N_{\ell},\theta\right)  \right]
.
\]

\bigskip

Define the objective function%
\[
\widehat{\mathcal{Q}}\left(  \xi\right)  =L\left\{  \widehat{Q}\left(
\gamma\left(  \tau\right)  +\frac{\xi}{\sqrt{L}};\widehat{\theta}\right)
-\widehat{Q}\left(  \gamma\left(  \tau\right)  ;\theta\right)  -\widehat{Q}%
_{\vartheta}^{\prime}\left(  \widehat{\theta}-\theta\right)  -\frac{\left(
\widehat{\theta}-\theta\right)  ^{\prime}\widehat{Q}_{\vartheta\vartheta
}\left(  \widehat{\theta}-\theta\right)  }{2}\right\}
\]
which is such that%
\[
\sqrt{L}\left(  \widehat{\gamma}\left(  \tau\right)  -\gamma\left(
\tau\right)  \right)  =\arg\min_{\xi}\widehat{\mathcal{Q}}\left(  \xi\right)
.
\]
For simplicity of notation, denote $\Psi_{I_{\ell}^{\ast}}\left(  \tau;Z_{\ell},N_{\ell},\theta\right) = \Psi_{I_{\ell}^{\ast}}$. Arguing as in Pollard (1991, p.192) yields, for each fixed $\xi$,
\begin{equation}\label{eq:quadapprox}
L\left\{  \widehat{Q}\left(  \gamma\left(  \tau\right)  +\frac{\xi}{\sqrt{L}%
};\theta\right)  -\widehat{Q}\left(  \gamma\left(  \tau\right)  ;\theta
\right)  \right\}  =\widehat{S}^{\prime}\xi+\frac{1}{2}\xi^{\prime}H\left(
\tau\right)  \xi+ \sum_{\ell=1}^{L}\left(\widetilde{R}_{\ell}\left(\xi \right) - \mathbb{E}\left[\widetilde{R}_{\ell}\left(\xi \right)\right]\right)
%o_{\mathbb{P}}\left(  1\right)  .
\end{equation}
where 
\begin{align*}
\widetilde{R}_{\ell}\left(\xi \right) &= \left\{\rho_{\Psi_{I_{\ell}^{\ast}}
}\left(W_{\ell}-X^{\prime}_{\ell}\left(\gamma\left(\tau\right) + \frac{\xi }{\sqrt{L}}\right)\right) - \rho_{\Psi_{I_{\ell}^{\ast}}
}\left(W_{\ell}-X^{\prime}_{\ell}\gamma\left(\tau\right)\right)\right\} \\
&- \left(\frac{1}{\sqrt{L}} X_{\ell}\left[  \mathbb{I}%
\left(  W_{\ell}\leq X_{\ell}^{\prime}\gamma\left(  \tau\right)  \right)
-\Psi_{I_{\ell}^{\ast}}  \right] \right)^{\prime}\xi
\end{align*}
$\sum_{\ell=1}^{L}\left(\widetilde{R}_{\ell}\left(\xi \right) - \mathbb{E}\left[\widetilde{R}_{\ell}\left(\xi \right)\right]\right)$ contributes only $o_{\mathbb{P}}\left(  1\right)$ to \eqref{eq:quadapprox}. To see this note that $\rho_{a}(b) = (a-\mathbb{I}(b<0))b = \int_{0}^{b}(a-\mathbb{I}(t<0))dt$ and denote $\delta_{\ell}(\xi) = X_{\ell}^{\prime}\xi/\sqrt{L}$ and $\widetilde{D}_{\ell}(\tau)=W_{\ell}-X_{\ell}^{\prime}\gamma(\tau)$ 
\begin{align*}
\widetilde{R}_{\ell}\left(\xi \right)  &= \rho_{\Psi_{I_{\ell}^{\ast}}}\left(\widetilde{D}_{\ell}(\tau)-\delta_{\ell}(\xi)\right) - \rho_{\Psi_{I_{\ell}^{\ast}}}\left(\widetilde{D}_{\ell}(\tau)\right) - \delta_{\ell}(\xi)\left[\mathbb{I}\left(\widetilde{D}_{\ell}(\tau) \leq 0 \right) - \Psi_{I_{\ell}^{\ast}}(\tau;\theta) \right] \\
&= \int_{0}^{\delta_{\ell}(\xi)} \left[\mathbb{I}\left(\widetilde{D}_{\ell}(\tau) \leq t\right) - \mathbb{I} \left(\widetilde{D}_{\ell}(\tau) \leq 0\right) \right] dt. 
\end{align*}
Using Cauchy-Schwarz inequality
\[
\widetilde{R}_{\ell}\left(\xi \right)^{2} \leq \left\vert{\delta_{\ell}(\xi)}\right\vert \left\vert{\int_{0}^{\delta_{\ell}(\xi)} \mathbb{I}\left(\left\vert \widetilde{D}_{\ell}\left(\tau\right) \right\vert \leq \left\vert t \right\vert\right) dt}\right\vert \leq \left\vert{\delta_{\ell}(\xi)}\right\vert {\int_{0}^{\left\vert \delta_{\ell}(\xi)\right\vert} \mathbb{I}\left(\left\vert \widetilde{D}_{\ell}\left(\tau;\right) \right\vert \leq \left\vert t \right\vert\right) dt}.
\]
Denote $\vert\vert f_{W}(\cdot \vert \cdot) \vert\vert_{\infty} \text{=} \sup_{w,x,z} \vert f_{W}(w \vert X,Z,I^{\ast},N) \vert$.
\begin{align*}
\mathbb{E}\left[\widetilde{R}^{2}\left(\xi\right) \vert X,Z,I^{\ast},N \right] &\leq \left\vert{\delta(\xi)}\right\vert {\int_{0}^{\vert \delta(\xi) \vert} \left\{\int \mathbb{I} \left(\vert w - X^{\prime}\gamma\left(\tau\right) \vert \leq \vert t \vert \right) f_{W}(w \vert X,Z,I^{\ast},N) dw \right\}} dt , \\
&\leq \vert\vert f_{W}(\cdot \vert \cdot) \vert\vert_{\infty} \left\vert{\delta(\xi)}\right\vert {\int_{0}^{\vert \delta(\xi) \vert} \left\{\int \mathbb{I} \left(\vert w - X^{\prime}\gamma\left(\tau\right) \vert \leq \vert t \vert \right) dw \right\}} dt , \\
&\leq \vert\vert f_{W}(\cdot \vert \cdot) \vert\vert_{\infty} \left\vert{\delta(\xi)}\right\vert {\int_{0}^{\vert \delta(\xi) \vert}2 \vert t \vert dt} = \vert\vert f_{W}(\cdot \vert \cdot) \vert\vert_{\infty} \left\vert{\delta(\xi)}\right\vert^3 \leq \frac{C \vert\vert X \vert\vert^{3}\vert\vert \xi \vert\vert^3}{L^{3/2}}.\\
\mathbb{E}\left[\widetilde{R}^{2}\left(\xi\right)\right] &= \mathbb{E}\left[\mathbb{E}\left[\widetilde{R}^{2}\left(\xi\right) \vert X,Z,I^{\ast},N \right]\right] \leq  \mathbb{E}\left[\frac{C \vert\vert X \vert\vert^{3}\vert\vert \xi \vert\vert^3}{L^{3/2}} \right] \leq \frac{C \vert\vert \xi \vert\vert^3}{L^{3/2}}.
\end{align*}
Due to cancellation of cross-product terms
\begin{align*}
\mathbb{E}\left[\left\vert \sum_{\ell=1}^{L}\left(\widetilde{R}_{\ell}\left(\xi\right) - \mathbb{E}\left[\widetilde{R}_{\ell}\left(\xi\right)\right] \right) \right\vert^2\right] &\leq \sum_{\ell=1}^{L} \mathbb{E}\left[\widetilde{R}_{\ell}^2 \left(\xi\right) \right] \leq \frac{C \vert\vert \xi \vert\vert^3}{\sqrt{L}} = o(1). 
\end{align*}

Lemma 2.4 in \cn{NeweyMcFadden1994} also gives%

\begin{align*}
&  \widehat{Q}\left(  \gamma\left(  \tau\right)  +\frac{\xi}{\sqrt{L}%
};\widehat{\theta}\right)  -\widehat{Q}\left(  \gamma\left(  \tau\right)
+\frac{\xi}{\sqrt{L}};\theta\right)  -\widehat{Q}_{\vartheta}^{\prime}\left(
\widehat{\theta}-\theta\right) \\
&  \quad=\left[  \int_{0}^{1}\left\{  \widehat{Q}_{\vartheta}^{\prime}\left(
\gamma\left(  \tau\right)  +\frac{\xi}{\sqrt{L}};\theta+t\left(
\widehat{\theta}-\theta\right)  \right)  -\widehat{Q}_{\vartheta}^{\prime
}\left(  \gamma\left(  \tau\right)  ;\theta\right)  \right\}  dt\right]
\left(  \widehat{\theta}-\theta\right) \\
&  \quad=\frac{\left(  \widehat{\theta}-\theta\right)  ^{\prime}%
\widehat{Q}_{\vartheta\vartheta}\left(  \widehat{\theta}-\theta\right)  }%
{2}+\left(  \widehat{\theta}-\theta\right)  ^{\prime}\widehat{Q}%
_{\vartheta\gamma}\frac{\xi}{\sqrt{L}}+o_{\mathbb{P}}\left(  \frac{1}%
{L}\right) \\
&  \quad=\frac{\left(  \widehat{\theta}-\theta\right)  ^{\prime}%
\widehat{Q}_{\vartheta\vartheta}\left(  \widehat{\theta}-\theta\right)  }%
{2}+\left(  \widehat{\theta}-\theta\right)  ^{\prime}D\left(  \tau\right)
\frac{\xi}{\sqrt{L}}+o_{\mathbb{P}}\left(  \frac{1}{L}\right)  .
\end{align*}
Hence, for each fixed $\xi$,%
\begin{align*}
\widehat{\mathcal{Q}}\left(  \xi\right)   &  =L\left\{  \widehat{Q}\left(
\gamma\left(  \tau\right)  +\frac{\xi}{\sqrt{L}};\widehat{\theta}\right)
-\widehat{Q}\left(  \gamma\left(  \tau\right)  +\frac{\xi}{\sqrt{L}}%
;\theta\right)  -\widehat{Q}_{\vartheta}^{\prime}\left(  \widehat{\theta
}-\theta\right)  -\frac{\left(  \widehat{\theta}-\theta\right)  ^{\prime
}\widehat{Q}_{\vartheta\vartheta}\left(  \widehat{\theta}-\theta\right)  }%
{2}\right\} \\
&  +L\left\{  \widehat{Q}\left(  \gamma\left(  \tau\right)  +\frac{\xi}%
{\sqrt{L}};\theta\right)  -\widehat{Q}\left(  \gamma\left(  \tau\right)
;\theta\right)  \right\} \\
&  =\left(  \widehat{S}+D\left(  \tau\right)  \sqrt{L}\left(  \widehat{\theta
}-\theta\right)  \right)  ^{\prime}\xi+\frac{1}{2}\xi^{\prime}H\left(
\tau\right)  \xi+o_{\mathbb{P}}\left(  1\right) \\
&  =\left(  \widehat{S}+D\left(  \tau\right)  \widehat{\Sigma}\right)
^{\prime}\xi+\frac{1}{2}\xi^{\prime}H\left(  \tau\right)  \xi+o_{\mathbb{P}%
}\left(  1\right)
\end{align*}
where the last line is from (\ref{FS}). Applying the convexity arguments in
\cn{Pollard1991} then gives, since $\sqrt{L}\left(  \widehat{\gamma}\left(
\tau\right)  -\gamma\left(  \tau\right)  \right)  =\arg\min_{\xi
}\widehat{\mathcal{Q}}\left(  \xi\right)  $,%
\[
\sqrt{L}\left(  \widehat{\gamma}\left(  \tau\right)  -\gamma\left(
\tau\right)  \right)  =-H\left(  \tau\right)  ^{-1}\left(  \widehat{S}%
+D\left(  \tau\right)  \widehat{\Sigma}\right)  +o_{\mathbb{P}}\left(
1\right)  .
\]
Then the joint asymptotic distribution of $\sqrt{L}\left(  \widehat{\gamma
}\left(  \tau\right)  -\gamma\left(  \tau\right)  \right)  $ and $\sqrt
{L}\left(  \widehat{\theta}-\theta\right)  $ is the one of $-H\left(
\tau\right)  ^{-1}\left(  \widehat{S}+D\left(  \tau\right)  \widehat{\Sigma
}\right)  $ and $\widehat{\Sigma}$ by (\ref{FS}), which by the CLT is a
centered normal with covariance matrix%
\[
\left[
\begin{array}
[c]{cc}%
\operatorname*{Var}\left(  H\left(  \tau\right)  ^{-1}\left(  \widehat{S}%
+D\left(  \tau\right)  \widehat{\Sigma}\right)  \right)  &
-\operatorname*{Cov}\left(  H\left(  \tau\right)  ^{-1}\left(  \widehat{S}%
+D\left(  \tau\right)  \widehat{\Sigma}\right)  ,\widehat{\Sigma}\right) \\
-\operatorname*{Cov}\left(  \widehat{\Sigma},H\left(  \tau\right)
^{-1}\left(  \widehat{S}+D\left(  \tau\right)  \widehat{\Sigma}\right)
\right)  & \mathcal{I}\left(  \theta\right)  ^{-1}%
\end{array}
\right]  ,
\]
which can be written as in the Theorem.$\hfill\square$

\pagebreak

\renewcommand{\theequation}{B.\arabic{equation}}
\renewcommand{\thesubsection}{B.\arabic{subsection}}
\setcounter{table}{0}
\renewcommand{\thetable}{B.\arabic{table}}
\setcounter{figure}{0}
\renewcommand{\thefigure}{B.\arabic{figure}}
\setcounter{theorem}{0}
\renewcommand{\thetheorem}{B.\arabic{theorem}}

\section*{Appendix B  - Specification analysis \label{Testsapp}}

\subsection{Power specification}

\paragraph{Asymptotic normality.} The next Proposition establishes that the normalization used for the $\widehat{\xi}_{p,q}$ in (\ref{Xipq}) ensures they all have an asymptotic standard normal distribution.

\begin{proposition}
	\label{Xilim}
	Suppose Assumption \ref{Ind} and (\ref{Abident}) hold, and that there exists some real numbers $0\leq\eta_{p,q}<\infty$ such that $\sup_{p,q} \left|\frac{L_{p,q}}{L_{Asy}}-\eta_{p,q}\right|=o_{\mathbb{P}} (1)$, where the supremum is over the finite support of the distribution of $(P,Q)$ and $pq \neq 0$.		
	
	Then, for any  type proportion $(p,q)$ with $pq \neq 0$, $\widehat{\xi}_{p,q}$ converges in distribution to a standard normal when the sample size grows.
\end{proposition}

Note that the $\widehat{\xi}_{p,q}$'s are asymptotically dependent due to the common estimated parameter $\widehat{\lambda}$. Although it would be possible to derive their joint asymptotic distribution, it may be difficult to use in practice as, for instance, using it for a Chi-square statistic may give a very large asymptotic variance matrix, $39 \times 39$ in the application, which may be difficult to estimate or to invert. Hence we prefer to use a maximum statistic in Section \ref{Asyspec}.

\subparagraph{Proof of Proposition \ref{Xilim}.} Recall 
\[
\widehat{\omega}_{p,q}
=
\frac{1}{L_{p,q}}
\sum_{\ell=1}^{L}
\mathbb{I}
\left(
\text{Mill wins in auction $\ell$, }
(P_{\ell},Q_{\ell}) = (p,q)
\right)
\]
and let $\omega_{p,q} (\widehat{\lambda}) = \frac{p}{p+\widehat{\lambda}q}$, so that $\widehat{\xi}_{p,q}=\sqrt{L_{p,q}} \left(\widehat{\omega}_{p,q}-\omega_{p,q} (\widehat{\lambda})\right)/\widehat{\sigma}_{p,q}$. Define also $IM_{\ell}=\mathbb{I}
\left(\text{Mill wins in auction $\ell$, } P_{\ell}Q_{\ell} \neq 0 \right)$. Standard expansions give,
$\sum_{\ell}^{Asy}$ standing for sum over asymmetric auctions,
\begin{eqnarray*}	
	\widehat{\lambda}-\lambda
	& = &
	-
	\frac{\frac{1}{L_{Asy}}
		\sum_{\ell}^{Asy}
		\left(
		IM_{\ell}
		-
		\frac{P_{\ell}}{P_{\ell}+\lambda Q_{\ell}}
		\right)}{\frac{1}{L_{Asy}}
		\sum_{\ell}^{Asy}
		\frac{P_{\ell}Q_{\ell}}{\left(P_{\ell}+\lambda Q_{\ell}\right)^2}}
	+
	o_{\mathbb{P}}
	\left(\frac{1}{\sqrt{L_{Asy}}}\right)
	\\
	& = &
	-\frac{
		\sum_{s,t}^{Asy} \frac{L_{s,t}}{L_{Asy}} 
		\left(\widehat{\omega}_{s,t}-\omega_{s,t}\right)
	}{
		\sum_{s,t}^{Asy} \frac{L_{s,t}}{L_{Asy}} \omega_{s,t}\left(1-\omega_{s,t}\right)
	}
	+
	o_{\mathbb{P}}
	\left(\frac{1}{\sqrt{L_{Asy}}}\right),
	\\
	\omega_{p,q} (\widehat{\lambda}) - \omega_{p,q}
	&= &
	-\frac{pq}{\left(p+q \widehat{\lambda}\right)^2}
	\left(\widehat{\lambda}-\lambda\right)
	+
	o\left(\widehat{\lambda}-\lambda\right)
	\\
	&= &
	\omega_{p,q} 
	\left(1-\omega_{p,q}\right)
	\frac{
		\sum_{s,t}^{Asy} \frac{L_{s,t}}{L_{Asy}} 
		\left(\widehat{\omega}_{s,t}-\omega_{s,t}\right)
	}{
		\sum_{s,t}^{Asy} \frac{L_{s,t}}{L_{Asy}} \omega_{s,t}\left(1-\omega_{s,t}\right)
	}
	+
	o_{\mathbb{P}}
	\left(\frac{1}{\sqrt{L_{Asy}}}\right).
\end{eqnarray*}
Note also that
\[
\max_{(p,q),(s,t)}
\left|
\frac{\sqrt{L_{p,q}}\sqrt{L_{s,t}}}{L_{Asy}}
-
\sqrt{\eta_{p,q}\eta_{s,t}}
\right|
=
o_{\mathbb{P}} (1).
\]
It follows that
\begin{align*}
\sqrt{L_{p,q}}
\left(
\omega_{p,q} (\widehat{\lambda}) 
- 
\widehat{\omega}_{p,q}
\right)
& =
\left(
\frac{\omega_{p,q} 
	\left(1-\omega_{p,q}\right)}{
	\sum_{s,t}^{Asy} \frac{L_{s,t}}{L_{Asy}} \omega_{s,t}\left(1-\omega_{s,t}\right)
}
-1
\right)
\sqrt{L_{p,q}}
\left(
\widehat{\omega}_{p,q}
-
\omega_{p,q}
\right)
\\
&
\quad
+
\frac{\omega_{p,q} 
	\left(1-\omega_{p,q}\right)}{
	\sum_{s,t}^{Asy} \frac{L_{s,t}}{L_{Asy}} \omega_{s,t}\left(1-\omega_{s,t}\right)
}
\sum_{(s,t)\neq(p,q)}^{Asy} \frac{\sqrt{L_{p,q}L_{s,t}}}{L_{Asy}} 
\sqrt{L_{s,t}}
\left(\widehat{\omega}_{s,t}-\omega_{s,t}\right)
+o_{\mathbb{P}} (1)
\\
&
\stackrel{d}{\rightarrow}
\mathcal{N} (0,\sigma^2_{p,q})
\end{align*}
where
\begin{align*}
\sigma^2_{p,q}
& =
\left(
\frac{\omega_{p,q} 
	\left(1-\omega_{p,q}\right)}{
	\sum_{s,t}^{Asy} \eta_{s,t} \omega_{s,t}\left(1-\omega_{s,t}\right)
}
-1
\right)^2
\omega_{p,q} 
\left(1-\omega_{p,q}\right)
\\
& \quad
+
\left(
\frac{\omega_{p,q} 
	\left(1-\omega_{p,q}\right)}{
	\sum_{s,t}^{Asy} \eta_{s,t} \omega_{s,t}\left(1-\omega_{s,t}\right)
}
\right)^2
\eta_{p,q}
\sum_{(s,t)\neq (p,q)} \eta_{s,t} \omega_{s,t}\left(1-\omega_{s,t}\right),
\end{align*}
observing that $\omega_{s,t}\left(1-\omega_{s,t}\right)=0$ for symmetric auctions. As
$\widehat{\sigma}^2_{p,q}= \sigma^2_{p,q}+o_{\mathbb{P}} (1)$, the Proposition is proven. \hfill $\Box$

\paragraph{Bootstrap procedure.} The $p$ values of Section \ref{Asyspec} in the main text and of the next paragraph are computed using the following standard pairwise bootstrap procedure, which is valid under the Assumptions of Proposition \ref{Xilim}. 

\begin{enumerate}
	\item[] \textit{Step 1.} Draw a bootstrap sample of winning bids auction covariates and types proportion  $\left\{W_{b,\ell}, X_{b,\ell}, P_{b,\ell}, Q_{b,\ell}, 1 \leq
	\ell \leq L\right\}$ with replacement from the realized values \\
	$\left\{W_{\ell}, X_{\ell}, P_{\ell}, Q_{\ell}, 1 \leq \ell \leq L \right\}$. Iterate for $b=1,\ldots,B=10,000$
	
	\item[] \textit{Step 2.} Compute  the bootstrapped $\widehat{\lambda}_b$, $L_{p,q,b}$, $\widehat{\omega}_{p,q,b}$, $\widehat{\sigma}^2_{p,q,b}$,
	\[
	\xi_{p,q,b}
	=
	\sqrt{L_{p,q,b}}\frac{\omega_{p,q} \left(\widehat{\lambda}_b\right) - \widehat{\omega}_{p,q,b}
		-
		\left(\omega_{p,q} \left(\widehat{\lambda}\right) - \widehat{\omega}_{p,q}\right)
	}{\widehat{\sigma}_{p,q,b}}
	\]
	and $\max |\widehat{\xi}|_{b}
	=
	\max_{(p,q): L_{p,q,b}>30,pq\neq 0}
	\left|\widehat{\xi}_{p,q,b} \right|$.
	
	\item[] \textit{Step 3.} The bootstrapped $p$-value for $\max |\widehat{\xi}|$ is given by
	\[
	\frac{1}{B} \sum_{b=1}^{B} \mathbb{I} \left( \max |\widehat{\xi}| \leq \max |\widehat{\xi}|_{b} \right).
	\]
	In the next paragraph, the bootstrapped $p$-value for $\widehat{\xi}_{p,q}$ is given by
	$
	\frac{1}{B} \sum_{b=1}^{B} \mathbb{I} \left(  |\widehat{\xi}_{p,q}| \leq |\widehat{\xi}_{p,q,b}| \right)
	$.
\end{enumerate}

\begin{figure}[!htbp]
	\caption{Empirical Cumulative Distribution Function of p-values }
	\begin{minipage}{0.8\linewidth}
		\includegraphics[width=\linewidth]{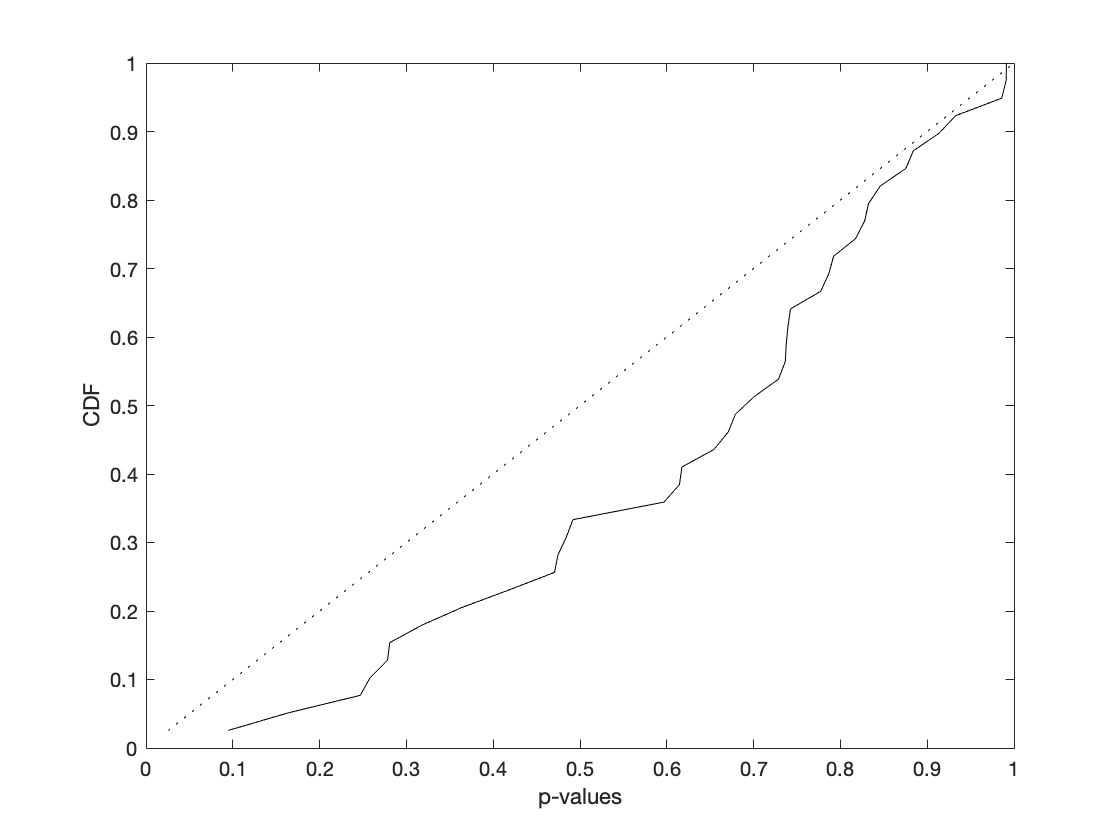}
	\end{minipage} \label{pvaluescdf}
\end{figure}

\paragraph{Additional figure and table.}
The next table gives the 39 selected type proportions $(p,q)$ with $L_{p,q}\geq 30$, the corresponding $L_{p,q}$, $|\widehat{\xi}_{p,q}|$ and associated individual $p$-values\footnote{The total number of auctions with $pq \neq 0$ and $L_{p,q}\geq 30$ is 3,290, about 44\% of the original sample.}. The $p$-values are quite high and only one $\left(p,q\right)=\left(2,5 \right)$ is lower than $10\%$. If the statistics $|\widehat{\xi}_{p,q}|$ were computed using the true $\lambda$ and $\sigma_{p,q}$ instead of estimations, the $p$-values would be, under $\text{H}_0^{Asy}$, independent draws from the uniform distribution. Figure \ref{pvaluescdf} shows indeed that the $p$-value cdf is close to the diagonal, as expected under the considered null.

\begin{table}[!htbp]
	\centering
	\caption{Asymmetry power specification analysis: results per type proportion}
	\begin{tabular}{cccccc}
		\hline\hline
		\multicolumn{2}{c}{Type's Proportion $(p,q)$} & $|\widehat{\xi}_{p,q}|$  & p-value & $L_{p,q}$ \\
		\hline
		1&	1&	0.61&	0.7005&	223\\
		1&	2&	0.33&	0.8324&	137\\
		1&	3&	0.18&	0.9136&	133\\
		1&	4&	0.41&	0.7773&	81\\
		1&	5&	0.32&	0.8279&	49\\
		1&	6&	1.76&	0.2581&	38\\
		2&	1&	0.01&	0.9910&	205\\
		2&	2&	0.22&	0.8840&	138\\
		2&	3&	1.17&	0.4707&	99\\
		2&	4&	0.79&	0.6174&	69\\
		2&	5&	2.90&	0.0947&	46\\
		3&	1&	0.48&	0.7376&	182\\
		3&	2&	1.06&	0.4918&	113\\
		3&	3&	0.71&	0.6544&	80\\
		3&	4&	1.99&	0.2469&	39\\
		3&	5&	0.79&	0.6146&	30\\
		4&	1&	0.42&	0.7425&	160\\
		4&	2&	1.01&	0.4842&	102\\
		4&	3&	0.02&	0.9859&	55\\
		4&	4&	0.35&	0.8174&	34\\
		5&	1&	1.09&	0.4183&	127\\
		5&	2&	0.52&	0.7287&	101\\
		5&	3&	0.65&	0.6790&	58\\
		6&	1&	1.26&	0.2808&	93\\
		6&	2&	1.46&	0.2783&	81\\
		6&	3&	0.36&	0.7922&	46\\
		7&	1&	0.84&	0.4746&	83\\
		7&	2&	0.11&	0.9327&	62\\
		7&	3&	0.46&	0.7396&	34\\
		7&	5&	0.44&	0.7866&	43\\
		8&	1&	0.20&	0.8755&	69\\
		8&	2&	0.55&	0.6709&	63\\
		8&	4&	1.46&	0.3639&	50\\
		9&	1&	1.61&	0.1638&	53\\
		9&	2&	0.01&	0.9913&	37\\
		9&	3&	0.28&	0.8460&	79\\
		10&	1&	1.31&	0.3181&	35\\
		10&	2&	0.44&	0.7365&	71\\
		11&	1&	0.59&	0.5968&	92\\
		\hline
		\hline
		\label{Xitesttab}
	\end{tabular}
\end{table}

\subsection{Power and parent distribution joint specification analysis}

\paragraph{Computation of $\widehat{G}(w,x|\widehat{\gamma}(\cdot),\widehat{\lambda})$ in (\ref{HatG0}) and truncation in (\ref{RW}).} The testing procedure uses a quantile level grid $\tau_i=i/100$, $i=1,\ldots,99$, and for each $X_{\ell}$ a value grid 
$v_j^{\ell} = X_{\ell}^{\prime} \widehat{\gamma}(0.01) + j X_{\ell}^{\prime} \left(\widehat{\gamma}(0.99)-\widehat{\gamma}(0.01) \right)/100$, $j=1,\ldots,100$.
The parent cdf $F(v_j^{\ell}|X_{\ell}) = \int_{0}^{1} \mathbb{I} \left[ X_{\ell}^{\prime} \gamma (t) \leq v\right] dt$ is estimated with the Riemann sum
\[
\widehat{F}(v_j^{\ell}|X_{\ell},\widehat{\gamma} (\cdot))
=
\frac{1}{100}
\sum_{i=1}^{99}
\mathbb{I}
\left[
X_{\ell}^{\prime} \widehat{\gamma} (\tau_i)
\leq
v_j^{\ell}
\right].
\]
The corresponding numerical computation of the winning bid quantile function is, using (\ref{Wapp}) and the rearrangement formula,
\begin{eqnarray}
\lefteqn{\widehat{W}(\tau_i|X_{\ell},P_{\ell},Q_{\ell},T_{\ell},\widehat{\gamma}(\cdot),\widehat{\lambda})
	=
	X_{\ell}^{\prime} \widehat{\gamma} (.01)
}
\nonumber
\\
&&
+
\frac{X_{\ell}^{\prime} \left(\widehat{\gamma}(0.99)-\widehat{\gamma}(0.01) \right)}{100}
\sum_{j=1}^{100}
\mathbb{I}
\left[
\psi
\left(
\left. 
\widehat{F}(v_j^{\ell}|X_{\ell},\widehat{\gamma} (\cdot))
\right|
P_{\ell},Q_{\ell},T_{\ell},\widehat{\gamma}(\cdot),\widehat{\lambda}
\right)
<
\tau_i
\right]
\label{Wi}.
\end{eqnarray}
The numerical approximation used for (\ref{HatG0}) is then 
\[
\hat{G}(W_{\ell},X_{\ell}|\widehat{\gamma}(\cdot),\widehat{\lambda})
=
\frac{1}{L}
\sum_{k=1}^{L}
\mathbb{I} (X_k \leq X_{\ell})
\frac{1}{100}
\sum_{i=1}^{99}
\mathbb{I}
\left[
\widehat{W}(\tau_i|X_{k},P_{k},Q_{k},T_{k},\widehat{\gamma}(\cdot),\widehat{\lambda})
\leq W_{\ell}
\right].
\]
The sum (\ref{RW}) defining the Rothe and Wied statistic $RW$ is restricted to auctions with transaction price $W_{\ell}$ in
$[X_{\ell}^{\prime} \widehat{\gamma} (.01),X_{\ell}^{\prime} \widehat{\gamma} (.99)]$ to avoid numerical errors at the boundaries, which may occur in view of the large  sample size of 7,462 observations.

\paragraph{The bootstrap procedure analysis.}
The two-step bootstrap procedure of \cn{RotheWied2013} is implemented as follows.
Preliminary to the bootstrap, we compute\\ $\widehat{W}(\tau_{i,B}|X_{\ell},P_{\ell},Q_{\ell},T_{\ell},\widehat{\gamma}(\cdot),\widehat{\lambda})$ over a larger quantile level grid
$\tau_{i,B}=i/1,000$, $i=1,\ldots,999$, using the value grid $v_j^{\ell} = X_{\ell}^{\prime} \widehat{\gamma}(0.001) + j X_{\ell}^{\prime} \left(\widehat{\gamma}(0.999)-\widehat{\gamma}(0.001) \right)/1,000$, $j=1,\ldots,1,00$. Then

\begin{enumerate}
	\item Draw with replacement $(X_{\ell,b}^{\star},P_{\ell,b}^{\star},Q_{\ell,b}^{\star},T_{\ell,b}^{\star})$, $\ell=1,\ldots,L$ from the initial auction sample. Use this bootstrap sample to estimate $\widehat{\lambda}^{\star}_b$;
	\item 
	Draw with replacement transaction price $W_{\ell,b}^{\star}$ from $\{\widehat{W}(\tau_{i,B}|X_{\ell,b}^{\star},P_{\ell,b}^{\star},Q_{\ell,b}^{\star},T_{\ell,b}^{\star},\widehat{\gamma}(\cdot),\widehat{\lambda}),i=1,\ldots,1,000\}$.
	Compute the statistic $RW_b^{\star}$ using the bootstrap sample $(W_{\ell,b}^{\star},X_{\ell,b}^{\star},P_{\ell,b}^{\star},Q_{\ell,b}^{\star},T_{\ell,b}^{\star})$, $\ell=1,\ldots,L$
\end{enumerate}
This is iterated for $b=1,\ldots,10,000$. The reported p-value is then $\frac{1}{10,000} \sum_{b=1}^{10,000} \mathbb{I} (RW \leq RW_b^{\star})$.

Note that the first-step of this procedure assumes that the conditional winner type distribution is correctly specified, as tested earlier. Such mispecifications can however be detected if they affect the functional form of the winning bid quantile function used in the second step. This procedure can be easily restricted to subsamples and modified to ensure that the bootstrapped number of each type proportion is identical to the one achieved in the initial sample, as done in Table \ref{RWtypetests} below.

\begin{table}[!htbp]
	\centering
	\caption{Rothe and Wied (2013) test p-value per type proportion}
	\begin{tabular}{cr|cc|cc}
		\hline\hline
		&			 & \multicolumn{2}{c|}{$\widehat{\gamma}(\cdot)$ whole sample} & \multicolumn{2}{c}{$\widehat{\gamma}(\cdot)$ w/o $N=12$}              
		\\
		(Mills, Loggers) &  $L_{p,q}$ & $RW(p,q)$  & p-value  & $RW(p,q)$  & p-value \\
		\hline
		(0,2)			&	202		& .0159	& .8418	& .0152	& .8704 \\
		(0,3)			&	159		& .0359	& .1796	& .0355	& .1938	\\
		(0,4)			&	120		& .0346	& .1489	& .0322	& .1881	\\
		(2,0)			&	715		& .0799	& .1350	& .0798	& .1617	\\
		(3,0)			&	557		& .1755	& .0058	& .1745	& .0105	\\
		(4,0)			&	438		& .1133	& .0094	& .1151	& .0129	\\
		(5,0)			&	304		& .1039	& .0099	& .0947 & .0173	\\
		(6,0)			&	239		& .0532	& .0782	& .0476	& .1219	\\
		(7,0)			&	196		& .0250	& .5063	& .0263	& .4787	\\
		(8,0)			&	127		& .0358	& .2713	& .0328	& .3422	\\
		(12,0)			&	160		& .1447	& .0041	& .1805	& .0012	\\
		(1,1)			&	223		& .0170	& .8639	& .0170	& .8677	\\
		(1,2)			&	137		& .0297	& .5030	& .0290	& .5272	\\
		(1,3)			&	133		& .0156	& .8849	& .0160	& .8724	\\
		(2,1)			&	205		& .0400	& .2148	& .0407	& .2202	\\
		(2,2)			&	138		& .0130	& .9579	& .0145	& .9234	\\
		(3,1)			&	182		& .0215	& .6125	& .0208	& .6505	\\
		(3,2)			&	113		& .0557	& .0918	& .0606	& .0737	\\
		(4,1)			&	160		& .0202	& .6149	& .0195	& .6521	\\
		(4,2)			&	102		& .0307	& .4199	& .0326	& .3820	\\
		(5,1)			&	127		& .0102	& .9846	& .0104	& .9837	\\
		(5,2)			&	101		& .0411	& .2750	& .0448	& .2290	\\
		Rest w/o $N=12$	&	2,175	& .0853	& .0257	& .1265	& .0123	\\
		$N=12$			&	449		& .2697	& .0000	& .3198	& .0000	\\
		\hline
		$\max_{N\neq 12} |RW(p,q)|$	&&  .1755 & .0255 & .1745 &	.0255 \\
		\hline\hline		
	\end{tabular}
	\label{RWtypetests}
\end{table}

\paragraph{Type proportion conditional specification analysis.}
The Rothe and Wied (2013) statistic $RW$, and to some extent the associated p-value, has a goodness of fit interpretation, as $0 \leq RW$ with a value of $0$ indicating that the model perfectly reproduces the sample distribution of $(W_{\ell},X_{\ell})$. Computing a statistic $RW(p,q)$ with empirical and model cdf's using subsamples given by a certain type proportion $(p,q)$ allows to analyze how well the model fits given mills and loggers participation, as done in  Table \ref{RWtypetests} where $\gamma (\cdot)$ and $\lambda$ are estimated over the whole sample or excluding the auction with twelve bidders. Table \ref{RWtypetests} computes such statistics $R_{p,q}$ for subsamples with a number $L_{p,q}$ of observations larger than $100$, the complement of these subsamples excluding auctions with twelve bidders, and a the remaining auctions with twelve bidders. 

\begin{figure}
	\centering
	\includegraphics[width=0.8\linewidth]{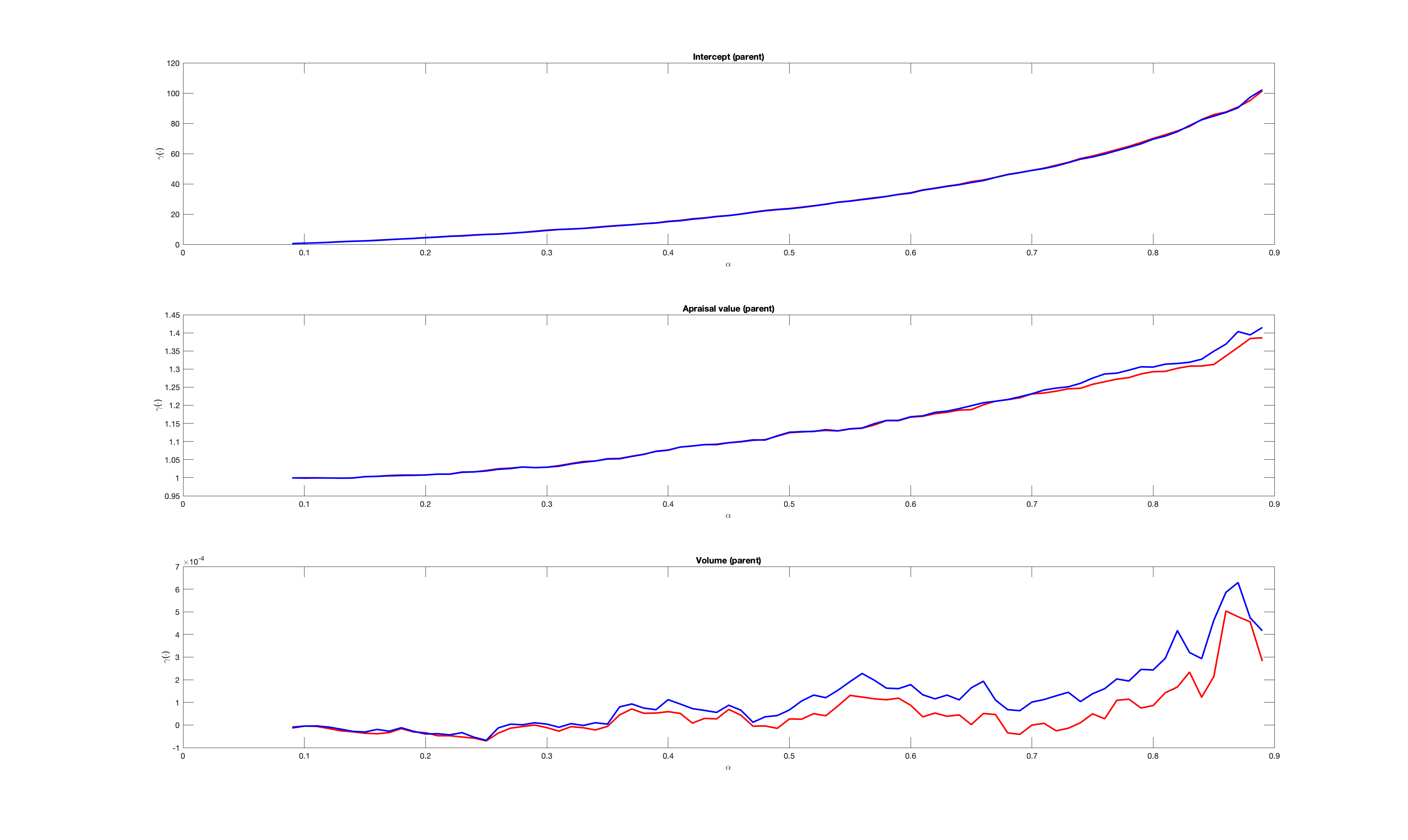}
	\caption{Parent quantile estimation with (in blue) or  without N=12 (in red): Intercept (top), appraisal value (middle), volume (bottom). }
	\label{fig:graphasgamman12}
\end{figure}

Table \ref{RWtypetests} shows that the model has a poor fit for auctions with twelve bidders, which may confirm that participation is mismeasured for these data as reported in \cn{AradillasGandhiQuint2013}. Otherwise the fit of the model seems quite good, except maybe for symmetric Mills auctions with $p=3,4,5$ and $q=0$. Note that estimating $\gamma (\cdot)$ and $\lambda$ excluding auctions with twelve bidders seems to improve the fit, giving p-values higher than $1 \%$ for these symmetric Mills auctions. Table \ref{RWtypetests} also reports the bootstrapped p-values of maximum $RW(p,q)$ statistics, which are smaller than the ones obtained in Table \ref{RWtests}, as expected since such maximum statistic is driven by the worst case, but still larger than $1\%$.

\paragraph{Estimation impact of auction with twelve bidders.} Excluding auctions with twelve bidders gives an estimated $\lambda$ at $0.7070$, which is virtually identical to the full sample one. The next figure reports the results for the estimation of the parent quantile coefficient $\gamma(\cdot)$. Only the volume slope function looks affected, but Figure \ref{QFcoef} suggests that the volume slope function may not be significant.

\pagebreak

\renewcommand{\theequation}{C.\arabic{equation}}
\renewcommand{\thesubsection}{C.\arabic{subsection}}
\setcounter{table}{0}
\renewcommand{\thetable}{C.\arabic{table}}

\section*{Appendix C - Additional tables} \label{tables sect}

This appendix displays tables that were commented on but not included in the empirical application. All tables are for median ascending auctions. The second column in all the three tables gives the corresponding estimates considering the methodology proposed in \cn{Gimenes2017} with symmetric bidders for $N = 2,3,\ldots, 12$. If the econometrician does not take asymmetry into account, the seller's expected revenue, $\Pi\left( r | X, N, V_{0}\right)$ is
\begin{align*}
\Pi\left( r | X, N, V_{0}\right) & = V_{0}r^{N} + R N r^{N-1}\left(1 - r \right) \\
& \quad \quad + N(N-1) \int_{r}^{1}V\left( t | X\right) t^{N-2} \left(1 - t \right)dt 
\end{align*}
and the optimal reserve price is $R_{\ast}=V(r_{\ast}|X)$ with $r_{\ast}=\arg \max_r \Pi\left( r | X, N, V_{0}\right)$. The seller value $V_0$ is set to $0$ and the private value quantile function is estimated as in 
\cn{Gimenes2017}. Note that this differs from Section \ref{Symmiss}, where the true asymmetric distribution was used to compute the revenue achieved using a optimal reserve price from the misspecified symmetric model.
Estimates taking into account asymmetry among the bidders, as discussed in this paper, are given on columns three to eight. They were computed using the expressions \ref{ER} and \ref{RP}. The proportion of bidder's types are given in parentheses following the rule $\left(\# Loggers, \# Mills \right)$.

\begin{landscape}
	\begin{table}[p!]
		\centering
		\caption{Seller Non Strategical Expected Revenue as a Function of the Proportion of Types }
		\resizebox{\textwidth}{!}{ 
			\begin{tabular}{cccccccc}
				\hline\hline
				& \cn{Gimenes2017}'s Approach & \multicolumn{6}{c}{Asymmetric Approach} \\
				&  &	(2,0) & (1,1) &	(0,2) &  &  &      \\				
				$N = 2$	& 54.76 & 48.02 &	52.46 &	57.65 	&  &  &   \\		
				& \small{[53.42, 56.68]}  &	\small{[46.37, 50.42]} &	\small{[51.07, 54.40]} &	\small{[56.20, 59.70]}	&  &  &   \\						
				& &	(3,0)  & (2,1)  &	(1,2)  & (0,3)	 &  &    \\				
				$N=3	$ & 70.69 & 63.59 &	67.30 &	71.18 &	75.04 	&  &   \\				
				& \small{[69.12, 72.83]} &	\small{[61.61, 66.03]} &	\small{[65.72, 69.44]} &	\small{[69.64, 73.31]} &	\small{[73.30, 77.47]}		&  &   \\		
				& &	(4,0) &	(3,1) &	(2,2) &	(1,3) &	(0,4) & \\
				$N=4$ &	82.99 &	74.82 &	78.23 &	81.63 &	84.96 &	88.17 & \\
				& \small{[81.36, 85.29]} &\small{[72.59, 77.49]} &	\small{[76.40, 80.58]} &	\small{[80.03, 83.85]} &	\small{[83.23, 87.32]} &	\small{[86.19, 90.83]} & \\		
				& &	(5,0) &	(4,1) &	(3,2) &	(2,3)&	(1,4) &	(0,5)	 \\	
				$N=5$ & 92.95 &	84.25 &	87.31 &	90.30 &	93.19 &	95.97 &	 98.64	 \\
				& \small{[91.05, 95.41]} &	\small{[81.82, 87.14]} & \small{[85.27, 89.86]} &	\small{[88.48, 92.62]} &	\small{[91.42, 95.58]} &	\small{[94.07, 98.54]} &	\small{[96.51, 101.45]}   \\				
				&	&	(6,0) &	(5,1) &	(4,2)  &  &	(1,5) &	(0,6) \\	
				$N=6$ &	101.14	& 92.31 &	95.02 &	97.65 & (\ldots) &	104.93	& 107.16	\\
				&	\small{[99.23, 103.68]} &	\small{[89.76, 95.32]}	 & \small{[92.84, 97.71]}	 &\small{ [95.68, 100.15]}	 &  &\small{ [102.90, 107.64]}	& \small{[104.87, 110.08]}	\\
				&	&	(7,0) &	(6,1) &	(5,2) 	 &  &	(1,6)	& (0,7) \\
				$N=7$ &	107.96 &	99.25 &	101.65 &	103.97 &	(\ldots) & 112.35 &	114.23 \\
				&	\small{[105.97, 110.64] }&	\small{[96.62, 102.31] }&	\small{[99.32, 104.41] }&\small{	[101.88, 106.60]} &	&	\small{[110.18, 115.17] }&\small{	[111.84, 117.24] } \\
				& &	(8,0) &	(7,1) &	(6,2) &	 &	(1,7) &	(0,8) \\
				$N=8$ & 113.72 &	105.28 &	107.41 &	109.47 & (\ldots)	  &	 118.57 &	120.17 \\
				& \small{[111.49, 116.55]}  &	\small{[102.57, 108.42]} &	\small{[104.99, 110.26]} &	\small{[107.26, 112.19]} & &	 	\small{[116.26, 121.51]} &	\small{[117.68, 123.34]}\\
				&  &	(9,0) &	(8,1) &	(7,2) & &	(1,8) &	(0,9) \\
				$N=9$ &	118.62 & 	110.56 &	112.46 &	114.30 & (\ldots) &	123.85 &	125.22 \\
				&	\small{[116.45, 121.56]} &	\small{[107.80, 113.69]}	 &\small{[110.01, 115.36] }&	\small{[112.01, 117.07]}	&  &	\small{[121.42, 126.98]} &\small{	[122.64, 128.47]} \\
				& &	(10,0) &	(9,1) &	(8,2) &	& 	(1,9) &	(0,10) \\
				$N=10$ & 122.83 &	115.22 &	116.93 &	118.56 & (\ldots)	 &	 128.36 &	129.55 \\
				& \small{[120.30, 125.84]}  &	\small{[112.45, 118.58]} &	\small{[114.23, 120.07]} &	\small{[115.96, 121.55]} & &	\small{[125.56, 131.63]} &	\small{[126.59, 132.95]}\\
				& &	(10,0) &	(9,1)	& (8,2)	&  & 	(1,9) &	(0,10) \\
				$N=11$ &	126.47 &	119.36 &	120.89 &	122.36 & (\ldots) &	132.25	 & 133.28 \\
				&	\small{[124.11, 129.56] }&	\small{[116.54, 122.48]}	&\small{ [118.31, 123.86] }&	\small{[120.01, 125.19]}	 &  &\small{ [129.56, 135.51]} &	\small{[130.48, 136.65] }\\
				&  &	(12,0) &	(11,1) &	(10,2) &	& 	(1,11) &	(0,12) \\
				$N=12$ & 129.62 &	123.05 &	124.43 &	125.76 & (\ldots)	 &	 135.62 &	136.52 \\
				& \small{[126.93, 132.78]}  &	\small{[120.23, 126.19]} &	\small{[121.91, 127.43]} &	\small{[123.32, 128.65]} & &	\small{[132.83, 138.94]} &	\small{[133.61, 139.96]}\\
				\hline
				\label{count_prop_ER_nonstr}
		\end{tabular}}
		\begin{flushleft}  \scriptsize  The 95\% confidence intervals for the quantile regression estimates were computed by resampling with replacement the ($X_{\ell},W_{\ell}$)-pair. As one goes from left to right, loggers are replaced by mills and the proportion of mills increases keeping the number of bidders fixed. In the vertical direction, as one goes from the top of the table to the bottom, the number of mills is kept fixed and loggers are included increasing the number of bidders. \end{flushleft}
	\end{table}
\end{landscape}

\begin{landscape}
	\begin{table}[p!]
		\centering
		\caption{Seller Strategical Expected Revenue as a Function of the Proportion of Types} 
		\resizebox{\textwidth}{!}{
			\begin{tabular}{cccccccc}
				\hline\hline
				& \cn{Gimenes2017}'s Approach & \multicolumn{6}{c}{Asymmetric Approach} \\
				&  &	(2,0) & (1,1) &	(0,2) &  &  &      \\				
				$N = 2$	& 73.77 & 69.65 &	73.1 &	76.44 	&  &  &   \\		
				& \small{[72.93, 74.97]}  &	\small{[68.64, 70.92]} &	\small{[72.31, 74.20]} &	\small{[75.45, 77.77]}	&  &  &   \\					
				& &	(3,0)  & (2,1)  &	(1,2)  & (0,3)	 &  &    \\				
				$N=3	$ & 83.37 & 77.97 &	81.07 &	84.06 &	86.95 	&  &   \\				
				& \small{[82.20, 84.97]} &	\small{[76.60, 79.67]} &	\small{[79.97, 82.54]} &	\small{[82.97, 85.65]} &	\small{[85.62, 88.73]}		&  &   \\		
				&	&	(4,0) &	(3,1) &	(2,2) &	(1,3) &	(0,4) & \\
				$N=4$ &	91.73 &	85.44 &	88.24 &	90.93 &	93.53 &	96.03 & \\
				&	\small{[90.46, 93.64]}	 &\small{[83.78, 87.48]} &	\small{[86.88, 90.02]} &	\small{[89.62, 92.71]}	 &\small{[92.15, 95.45]} &	\small{[94.41, 98.17]} & \\
				& &	(5,0) &	(4,1) &	(3,2) &	(2,3)&	(1,4) &	(0,5)	 \\	
				$N=5$ & 99.03 &	92.16 &	94.69 &	97.12 &	99.46 &	101.72 &	 103.9	 \\
				& \small{[97.37, 101.24]} &	\small{[90.27, 94.45]} &	\small{[93.13, 96.71]} &	\small{[95.59, 99.09]} &	\small{[97.99, 101.56]} &	\small{[100.08, 103.95]} &	\small{[102.06, 106.30]}	  \\				
				& &		(6,0) &	(5,1) &	(4,2) &  &	(1,5) &	(0,6)	\\
				$N=6$ &	105.4 &	98.22	 &100.51 &	102.71 & (\ldots)
				&	108.83 &	110.73	\\
				&	\small{[103.80, 107.78]} &	\small{[96.15, 100.73]}	& \small{[98.74, 102.75]}	&\small{[101.03, 104.89]}	 & &\small{[106.98, 111.29]}&	\small{[108, 68, 113.36]}	\\
				& &	(7,0) &	(6,1) &	(5,2)	&  &	(1,6) &	(0,7) \\
				$N=7$	&110.98 &	103.69 &	105.76 &	107.75 & (\ldots) &	115.02	 &116.67 \\
				&	\small{[109.23, 113.54]} &	\small{[101.50, 106.36]} &	\small{[103.83, 108.15]}	 &\small{[105.91, 110.11]}	&  &\small{	[112.96, 117.68]} &	\small{[114.48, 119.45]} \\
				& &	(8,0) &	(7,1) &	(6,2) &	 &	(1,7) &	(0,8) \\
				$N=8$ & 115.88 &	108.63 &	110.51 &	112.32 & (\ldots)	  &	 120.41 &	121.86 \\
				& \small{[113.72, 118.60]}  &	\small{[106.32, 111.36]} &	\small{[108.47, 113.02]} &	\small{[110.35, 114.83]} & &	 	\small{[118.21, 123.23]} &	\small{[119.54, 124.84]}\\
				&	&	(9,0) &	(8,1)&	(7,2)& &	(1,8) &	(0,9) \\
				$N=9$ &	120.17&	113.11&	114.81&	116.46 & (\ldots) &	125.13&	126.39 \\
				&	\small{[118.16, 122.99]}	&\small{[110.68, 115.90]}&	\small{[112.66, 117.45]}	&\small{[114.40, 119.09]}	 &  &	\small{[122.79, 128.11]}	&\small{[123.94, 129.52]} \\
				& &	(10,0) &	(9,1) &	(8,2) &	& 	(1,9) &	(0,10) \\
				$N=10$ & 123.95 &	117.16 &	118.72 &	120.21 & (\ldots)	 &	 129.26 &	130.37 \\
				& \small{[121.56, 126.89]}  &	\small{[114.70, 119.99]} &	\small{[116.47, 121.48]} &	\small{[118.09, 122.94]} & &	\small{[126.79, 132.35]} &	\small{[127.76, 133.61]}\\
				& &		(11,0) &	(10,1) &	(9,2) &  & (1,10)	 & (0,11) \\
				$N=11$	& 127.27&	120.85&	122.26&	123.62& (\ldots)	 & 132.88	&133.86 \\
				&	\small{[125.06, 130.29]}	&\small{[118.33, 123.75]}	&\small{[119.93, 125.11]}&	\small{[121.42, 126.43]} &  & \small{[130.29, 136.05]}&	\small{[131.14, 137.17]}\\
				&  &	(12,0) &	(11,1) &	(10,2) &	& 	(1,11) &	(0,12) \\
				$N=12$ & 130.2 &	124.2 &	125.48 &	126.72 & (\ldots)	 &	 136.06 &	136.92 \\
				& \small{[127.63, 133.36]}  &	\small{[121.63, 127.20]} &	\small{[123.11, 128.40]} &	\small{[124.40, 129.60]} & &	\small{[133.35, 139.36]} &	\small{[134.11, 140.35]}\\
				\hline
				\label{count_prop_ER}
		\end{tabular}}
		\begin{flushleft}  \scriptsize  The 95\% confidence intervals for the quantile regression estimates were computed by resampling with replacement the ($X_{\ell},W_{\ell}$)-pair. As one goes from left to right, loggers are replaced by mills and the proportion of mills increases keeping the number of bidders fixed. In the vertical direction, as one goes from the top of the table to the bottom, the number of mills is kept fixed and loggers are included increasing the number of bidders. \end{flushleft}
	\end{table}
\end{landscape}

\begin{landscape}
	\begin{table}[p!]
		\centering
		\caption{Optimal Reserve Price as a Function of the Proportion of Types} 
		\resizebox{\textwidth}{!}{
			\begin{tabular}{cccccccc}
				\hline\hline
				& \cn{Gimenes2017}'s Approach & \multicolumn{6}{c}{Asymmetric Approach} \\
				&  &	(2,0) & (1,1) &	(0,2) &  &  &     \\				
				$N = 2$ & 107.85	 & 104.65 &	106.88 &	 111.85 	&  &  &  \\		
				& \small{[102.73, 116.36]}  &	\small{[102.17, 114.11]} &	\small{[103.33, 116.24]} &	\small{[103.82, 117.47]}	&  &  &   \\			
				& &	(3,0)  & (2,1)  &	(1,2)  & (0,3)	 &  &     \\	
				$N=3$ &	107.85 &	104.65 &	104.65 &	106.88  &	111.85 & & \\
				&	\small{[102.80, 116.36]} &	\small{[102.16, 114.11]} &	\small{[103.0, 115.71]} &	\small{[103.58, 116.70]} &	\small{[103.90, 117.47]}	 &	& \\
				&	&	(4,0) &	(3,1) &	(2,2) &	(1,3) &	(0,4) & \\			
				$N=4$ & 107.85 & 104.65 &	104.65 &	106.88 & 106.88 &	111.85 	 &   \\			
				&  \small{[102.80, 116.36]}  &	\small{[102.17, 114.11]} &	\small{[102.87, 115.57]} &	\small{[103.33, 116.31]} &	\small{[103.94, 117.76]}		&  &   \\	
				& &	(5,0) &	(4,1) &	(3,2) &	(2,3)&	(1,4) &	(0,5) \\	
				$N=5$ & 107.85 &	104.65 &	104.65 &104.65 &	106.88 &	111.85 &	 111.85	 \\
				&  \small{[102.80, 116.38]} &	\small{[102.17, 114.11]} &	\small{[102.81, 115.57]} &	\small{[103.12, 115.99]} &	\small{[103.54, 116.70]} &	\small{[103.66, 117.13]} &	\small{[103.97, 117.75]}	   \\	
				& &		(6,0) &	(5,1) &	(4,2) &  &	(1,5) &	(0,6)	\\
				$N=6$ &	107.85 &	104.65 &	104.65	& 104.65 & (\ldots)
				&	111.85 &	111.85	\\
				&	\small{[102.88, 116.38]}	& \small{[102.17, 114.11]} &	\small{[102.79, 115.57]}	& \small{[103.0, 115.92]}	 & &\small{[103.72, 117.41]}	&\small{[103.97, 117.76]}	\\
				& &	(7,0) &	(6,1) &	(5,2)	&  &	(1,6) &	(0,7) \\
				$N=7$	&107.85 &	104.65	 &104.65&	104.65 & (\ldots) &	111.85 &	111.85 \\
				&	\small{[102.88, 116.38]}	& \small{[102.19, 114.15]}&	\small{[102.59, 115.41]} &	\small{[102.98, 115.71]}	&  &\small{	[103.77, 117.62]} &	\small{[103.97, 117.77]} \\	
				& &	(8,0) &	(7,1) &	(6,2) &	& 	(1,7) &	(0,8) \\
				$N=8$ & 107.85 &	104.65 &	104.65 &	104.65 & (\ldots)	 &	 111.85 &	111.85 \\
				&  \small{[102.88, 116.41]}  &	\small{[102.19, 114.26]} &	\small{[102.51, 115.41]} &	\small{[102.95, 115.71]} & & 	\small{[103.77, 117.61]} &	\small{[103.97, 117.77]}\\
				&	&	(9,0) &	(8,1)&	(7,2)& &	(1,8) &	(0,9) \\
				$N=9$ &	107.85	 & 104.65 &	104.65 &	104.65 & (\ldots) &	111.85	 & 111.85 \\
				&	\small{[102.85, 116.41]} &	\small{[102.19, 114.64]}	 &\small{[102.51, 115.41]} &	\small{[102.93, 115.71]}	 &  &	\small{[103.77, 117.62]}	& \small{[103.90, 117.77]} \\
				& &	(10,0) &	(9,1) &	(8,2) &	&	(1,9) &	(0,10) \\
				$N=10$ & 107.85 &	104.65 &	104.65 &	104.65 & (\ldots)	 &	 111.85 &	111.85\\
				&  \small{[102.85, 116.41]}  &	\small{[102.19, 114.64]} &	\small{[102.50, 115.41]} &	\small{[102.81, 115.68]} & & 	\small{[103.77, 117.61]} &	\small{[103.90, 117.75]}\\
				& &		(11,0) &	(10,1) &	(9,2) &  & (1,10)	 & (0,11) \\
				$N=11$	& 107.85 & 104.65 &	104.65 &	104.65 & (\ldots)	 & 111.85	 &111.85 \\
				&	\small{[102.80, 116.41]} &	\small{[102.19, 114.73]}	 &\small{[102.48, 115.41]} &	\small{[102.80, 115.68]} &  & \small{[103.77, 117.62]} &	\small{[103.82, 117.75]} \\
				& &	(12,0) &	(11,1) &	(10,2) &	& 	(1,11) &	(0,12) \\
				$N=12$ & 107.85 &	104.65 &	104.65 &	104.65 & (\ldots)	 &		 111.85 &	111.85 \\
				&  \small{[102.73, 116.41]}  &	\small{[102.19, 114.73]} &	\small{[102.45, 115.41]} &	\small{[102.79, 115.49]} & &	\small{[103.77, 117.61]} &	\small{[103.77, 117.75]}\\
				\hline
				\label{count_prop_ORP}
		\end{tabular}}
		\begin{flushleft}  \scriptsize  The 95\% confidence intervals for the quantile regression estimates were computed by resampling with replacement the ($X_{\ell},W_{\ell}$)-pair. As one goes from left to right, loggers are replaced by mills and the proportion of mills increases keeping the number of bidders fixed. In the vertical direction, as one goes from the top of the table to the bottom, the number of mills is kept fixed and loggers are included increasing the number of bidders. \end{flushleft}
	\end{table}
\end{landscape}

\end{document}